\documentclass{article}

\usepackage{arxiv}

\usepackage[utf8]{inputenc} 
\usepackage[T1]{fontenc}    
\usepackage{hyperref}       
\usepackage{url}            
\usepackage{booktabs}       
\usepackage{amsfonts}       
\usepackage{nicefrac}       
\usepackage{microtype}      
\usepackage{lipsum}		
\usepackage{graphicx}
\usepackage[square,numbers]{natbib}
\usepackage{doi}

\usepackage{amsmath}
\usepackage{amssymb}
\usepackage{amsthm}

\newtheorem{theorem}{Theorem}[section]

\usepackage{tikz}
\usepackage{pgfplots}
\pgfplotsset{compat=1.18}

\usepackage{caption}
\usepackage{subcaption}
\usepackage{threeparttable}

\usepackage{algorithm}
\usepackage{algpseudocode}

\title{Subspace Acceleration for Efficient Nonlinear Water
Wave Simulation}

\author{ 
    \href{https://orcid.org/0009-0008-4400-7454}{\includegraphics[scale=0.06]{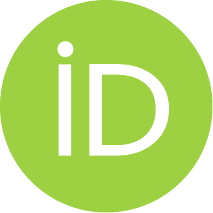}\hspace{1mm}Rasmus Kleist Hørlyck Sørensen} \\
	Department of Applied Mathematics and Computer Science \\
	Technical University of Denmark \\
	2800 Kongens Lyngby \\
	\texttt{rkhso@dtu.dk} \\
	\And
	\href{https://orcid.org/0000-0002-2923-0374}{\includegraphics[scale=0.06]{orcid.pdf}\hspace{1mm}Margherita Guido} \\
	Institute of Mathematics, Ecole Polytechnique \\
    Swiss Plasma Center \\
	Fédérale de Lausanne \\
	Vaud, Switzerland \\
	\texttt{margherita.guido@epfl.ch} \\
    \And
    \href{https://orcid.org/0000-0001-8626-1575}{\includegraphics[scale=0.06]{orcid.pdf}\hspace{1mm}Allan Peter Engsig-Karup} \\
	Department of Applied Mathematics and Computer Science \\
	Technical University of Denmark \\
	2800 Kongens Lyngby \\
	\texttt{apek@dtu.dk} \\
	\And
	\href{https://orcid.org/0000-0003-3369-2958}{\includegraphics[scale=0.06]{orcid.pdf}\hspace{1mm}Daniel Kressner} \\
	Institute of Mathematics, Ecole Polytechnique \\
	Fédérale de Lausanne \\
	Vaud, Switzerland \\
	\texttt{daniel.kressner@epfl.ch} \\
}

\hypersetup{
pdftitle={Subspace Acceleration for Efficient Nonlinear Water
Wave Simulation},
pdfsubject={},
pdfauthor={Rasmus K.H.~Sørensen},
pdfkeywords={Subspace Acceleration, Iterative Methods, Nonlinear Water Waves, Free Surface Incompressible Navier-Stokes, High-order finite Difference Method},
}

\begin{document}
\maketitle

\begin{abstract}

Efficient simulation of nonlinear and dispersive free-surface flows governed by the incompressible Navier-Stokes equations remains a central challenge in ocean and coastal engineering. The computational bottleneck arises from solving a time-dependent discretized Poisson problem at every time step to enforce divergence free flow. This is crucial to ensure conservation of mass and requires solving long sequences of time-dependent linear systems typically using iterative methods, such as the preconditioned Krylov subspace methods.
In this work, we investigate new subspace acceleration techniques for improving initial guesses to reduce the number of iterations required by iterative solvers, with a focus on nonlinear wave propagation problems. We extend the original subspace acceleration method by incorporating the complete history of previous solutions through an exponentially weighted formulation. This approach eliminates the need for repeated sketching and orthonormalization, resulting in a more efficient and scalable strategy to generate better initial guesses. Our method is implemented within a high-order finite-difference framework using a method-of-lines formulation and a low-storage Runge-Kutta time integration scheme. We demonstrate that subspace acceleration significantly reduces the number of GMRES iterations when solving the Poisson equation in nonlinear water wave simulations. Performance is evaluated on two benchmark problems: nonlinear stream function wave propagation and harmonic wave generation over a submerged bar. In both cases, the new approach achieves substantial improvements in computational efficiency without compromising accuracy. Although demonstrated using high-order finite difference methods, the technique is discretization independent and broadly applicable to incompressible free-surface flow solvers.

\end{abstract}


\clearpage

\section{Introduction}

Accurate modeling of nonlinear and dispersive free surface flows is important across applications in ocean and offshore engineering, e.g., to estimate regional sea states through simulations at increasing resolution and scale \cite{Soomere2023}. It is therefore of practical importance to accurately and robustly perform numerical simulations of nonlinear and dispersive wave propagation, especially under the influence of varying seabed topography. Hence, it is especially important to reduce runtime requirements for practical simulation scenarios that require large-scale regional or high-fidelity simulations, as described by \cite{Sriram2021} and \cite{Windt2018}, or in many-query studies that are common in parameterized structural design iterations, e.g., data set generation for offshore wind applications addressed by \cite{Pierella2021}. The objective of simulations is to accurately and cost-effectively model complex systems at scale, and these can greatly benefit from a reduced time-to-solution to enhance efficiency and scalability.

Incompressible Navier-Stokes equations are widely used in computational fluid dynamics to model nonlinear water waves in combination with a free surface boundary condition to capture the evolution of the fluid interface \cite{Engsig-Karup2013}. This framework provides an accurate description of the flow, taking into account rotational and viscous effects. To facilitate efficient computations in a fixed spatial domain, a \( \sigma \)-coordinate transformation is often introduced; for example, see the early work by \cite{Li2001}. This maps the vertical coordinate from the seabed to the still-water level, transforming the time-dependent fluid domain into a reference domain with fixed geometry. For incompressible flow solvers, a velocity-pressure coupling is needed to ensure mass conservation, and this is achieved by solving a Poisson problem for pressure that results in a linear system of equations. In time-varying domains that arise in free-surface incompressible Navier-Stokes equations, such a linear system becomes time-dependent, which can be formulated compactly as
\begin{equation}
    A(\tau) {\bf p}(\tau) = {\bf b}(\tau), \quad A\in\mathbb{R}^{n\times n},\quad {\bf p}, {\bf b}\in\mathbb{R}^n. 
\end{equation}
Solving linear systems of this form is common for the solution of time-dependent partial differential equations and leads to a sequence of linear systems that must be solved at every time step of the numerical simulation. For example, in recent work, Engsig-Karup et al.~\cite{engsig-karup2024} developed a high-order finite-difference scheme to solve the free surface incompressible Navier-Stokes equations. Their method uses a recently formulated high-order temporal integration and enforces incompressibility at each Runge-Kutta stage by solving a mixed-stage Poisson problem for the dynamic pressure to ensure proper velocity-pressure coupling \cite{Melander2025}. This type of advanced numerical model is representative of PDE solvers that include a time-dependent linear system and is, therefore, considered in this work.

Solving the sequence of mixed-stage Poisson problems for pressure at every time step is crucial to ensure that the incompressibility constraint is satisfied and that the numerical scheme conserves mass. However, as pointed out by Engsig-Karup et al.~\cite{engsig-karup2009, engsig-karup2014} in related work on fully nonlinear potential flow (FNPF) models, solving the linear system that results from the discretization procedure is the main computational bottleneck in the simulation of nonlinear water waves. This prompts the use of iterative solvers and efficient preconditioners designed for large-scale simulations using efficient iterative solvers, where the size of the system \( n \) may be counted in billions of degrees of freedom, as in the work of Glimberg et al.~\cite{Glimberg2019}. Other works seek to identify ways to achieve the attainable accuracy limited by numerical discretization errors to  devise early stopping criteria \cite{engsig-karup2014}; this is outside the scope of this work.

A variety of iterative solvers and preconditioning techniques have been proposed to accelerate convergence for these types of elliptic Poisson-type problems. It is common to design iterative solver strategies based on Krylov-based techniques, such as the conjugate gradient or GMRES methods, which are designed to guaranty convergence for symmetric and non-symmetric linear systems, respectively. Even short recurrence defect correction methods \cite{engsig-karup2014} may be useful since this method supports lowering memory requirements and minimizes global communications in massively parallel implementations with constrained memory space, such as graphics processing units (GPUs)~\cite{Glimberg2019}. Iterative solvers often exhibit slow convergence and stagnation without preconditioning that is designed to solve the linear system at a reduced cost, for example, by using $\mathcal{O}(n)$-scalable multigrid techniques. Also, it is possible to improve the efficiency of iterative solver strategies by providing good initial guesses. Such techniques often use the last solution computed for the linear system, extrapolation techniques, or projection techniques, as assessed by Austin et al.~\cite{austin2021}, where the focus is on a linear system in which the system matrix $A$ does not change over time. Related early work on project techniques for providing good initial guesses in incompressible flow simulations was done by Fischer~\cite{Fischer1998} and in transient field simulations by Clements et al.~\cite{clemens2004}.

Engsig-Karup et al.~\cite{engsig-karup2014} systematically analyzed  preconditioned defect correction and preconditioners for efficient, scalable, and low-storage iterative solutions of high-order accurate non-symmetric discretizations of the \( \sigma \)-transformed Laplace problem. In recent work, Melander et al.~\cite{melander2024} investigated multigrid preconditioners for a high-order hybrid spectral incompressible Navier-Stokes model for non-linear water wave simulation. Multigrid methods are particularly effective on structured meshes, where coarse grids and transfer operators can be defined naturally. Their performance and scalability have made them a popular choice for free-surface potential flow solvers and high-order Navier-Stokes solvers; see, for example, Li et al.~\cite{Li2001} and Engsig-Karup et al.~\cite{engsig-karup2009}. Recent work by Melander et al.\cite{melander2024} shows promising results for geometric multigrid preconditioners applied to high-order spectral discretizations of the mixed-stage Poisson problem that arises in free-surface incompressible Navier-Stokes solvers. It is also possible to achieve significant acceleration via parallel implementations of nonlinear and dispersive free surface solvers, e.g., using a parallel GPU implementation of a FNPF model \cite{Engsig-Karup2013}, a parallel CPU implementation of an FNPF \cite{visbech2025}, or even parallel-in-time techniques pioneered by Lions et al.~\cite{lions2001}. These techniques were first explored for accelerated water wave simulation by Engsig-Karup et al.~\cite{Engsig-Karup2013} using a GPU-accelerated high-order finite-difference FNPF solver \cite{engsig-karup2012} and in recent work by Poirier et al.~\cite{poirier2024} for a BEM solver. 

An alternative strategy for accelerating iterative solvers is subspace acceleration, introduced by Guido et al.~\cite{guido2024}. This method exploits the smooth temporal evolution of the solution to time-dependent linear systems to generate improved initial guesses for iterative solvers. This can be done explicitly through extrapolation methods or implicitly via projection-based techniques; see Austin et al. \cite{austin2021} for a review of these methods with application to incompressible flow. In the subspace acceleration framework, a subspace of the vector space spanned by previously computed solutions to  linear systems is constructed using techniques from randomized linear algebra\cite{murray2023}. By solving the new system restricted to this subspace in the least-squares sense, we obtain an initial guess that is typically much closer to the true solution than the previous solution.
The subspace acceleration technique is a natural generalization of the projection-based technique introduced by Fischer \cite{Fischer1998} to handle system matrices that vary in time.

\subsection{Contributions}
In this work, we apply and extend the subspace acceleration technique to the simulation of nonlinear free-surface flows governed by the incompressible Navier-Stokes equations. We focus on accelerating the solution of sequences of mixed-stage time-dependent Poisson problems that arise in the high-order finite-difference discretization proposed by Melander et al.~\cite{Melander2025}. Specifically, we extend the original subspace acceleration method proposed by Guido et al. \cite{guido2024} by proposing a computationally and memory efficient variant that uses the complete history of previous solutions. This eliminates the need for repeated sketching and orthonormalization, enabling faster and more scalable subspace construction.

We demonstrate the effectiveness of our subspace acceleration approach in the context of a high-order finite-difference discretization using a method-of-lines formulation and a low-storage explicit Runge-Kutta time integrator. Performance is evaluated on two standard benchmark problems: propagating nonlinear stream function waves and a classical test of harmonic wave generation over a submerged bar. In both cases, the proposed method achieves substantial reductions in the number of GMRES iterations per time step without compromising simulation accuracy. Although demonstrated here using a specific discretization framework, our proposed approach is discretization independent and is broadly applicable to time-dependent incompressible flow solvers.

\section{Governing Equations} \label{sec:navier:stokes}

\begin{figure}[!h]
    \centering
    \begin{tikzpicture}[scale=1.8]

        \def\w{4}
        \def\h{2}
        \draw[->] (0,0)--(1.125*\w,0) node[right]{\( x \)};
        \draw[->] (0,-1.25*\h)--(0,0.25*\h) node[above]{\( z \)};
        \filldraw (0,0) circle (1pt);
    
        \draw[domain=0:\w, smooth, variable=\x,thick] plot ({\x}, {0.2*\h*sin(180*\w/4*\x)});
    
        \draw[thick] (0,-\h) -- ({2*\w/16},-\h) -- ({3*\w/16},-{3*\h/4}) -- ({7*\w/16},-{3*\h/4}) -- ({8*\w/16},-\h) -- (\w,-\h);
    
        \draw[thick] (0,0) -- (0,-\h);
        \draw[thick] (\w,-\h) -- (\w,0);
    
        \draw[<->] ({(0.5+0.125)*\w},0) -- ({(0.5+0.125)*\w},{0.2*\h}) node[anchor=south]{\( \eta \)};
        \draw[<->] ({(0.5+0.125)*\w},0) -- ({(0.5+0.125)*\w},{-\h}) node[anchor=north]{\( h \)};
    
        \draw ({\w / 2},{-\h / 2}) node {\( \Omega_t \)};
        \draw (0,0) node[anchor=east] {\( z=0 \)};
        \draw (0,-\h) node[anchor=east] {\( z=-h \)};
        \draw (\w,-\h) node[anchor=north] {\( x=L_x \)};
        
    \end{tikzpicture}
    \caption{The domain \( \Omega_t \) at time \( t \) with free surface \( \eta(x,t) \) and still-water depth \( h(x) \) in two spatial dimensions.}
    \label{fig:domain}
\end{figure}
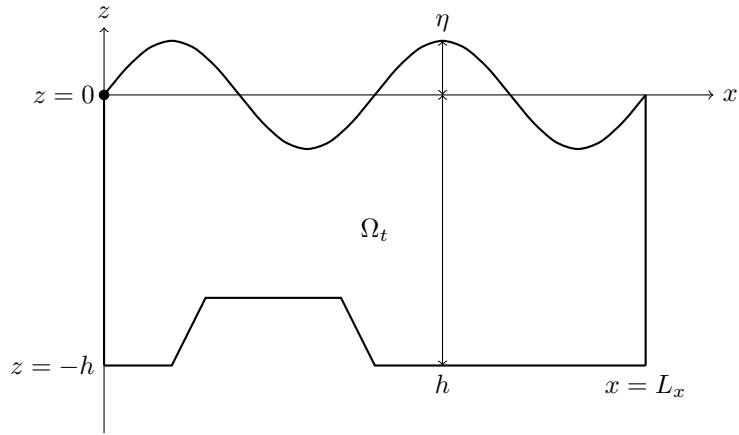

To model nonlinear and dispersive water waves, we consider a time-dependent spatial domain \( \Omega_t \), parameterized by \( t \in (0,T) \) for some \( T > 0 \). At each time \( t \), the spatial domain \( \Omega_t\) represents the volume of the fluid, which is bounded from below by the depth at still water  \( h \) and the elevation of the free surface \( \eta \) from above, as illustrated in Figure \ref{fig:domain}. 

The velocity of the fluid within the space-time cylinder \( \Omega_t\times(0,T) \) is described by a vector field \( \mathbf{u}:\Omega_t\times(0,T)\to\mathbb{R}^3 \) with \( \mathbf{u}=(u,v,w) \) and a scalar field \( p:\Omega_t\times(0,T)\to\mathbb{R} \). The velocity field \( \mathbf{u} \) and the pressure field \( p \) are coupled by the \emph{incompressible Navier-Stokes equations} with uniform viscosity:
\begin{equation} \label{eq:navier:stokes}
    \rho\frac{\partial \mathbf{u}}{\partial t} = \mu\nabla^2\mathbf{u} - \nabla p + \mathbf{F} - \rho(\mathbf{u} \cdot \nabla)\mathbf{u}, \quad\forall\mathbf{x}\in\Omega_t, \forall t\in(0,T), 
\end{equation}
where \( \rho \) is the density of the fluid, \( \mu \) is the dynamic viscosity, and \( \mathbf{F} \) represents external forces such as gravity. This equation expresses the conservation of momentum. For incompressible fluid flow, the density \( \rho \) remains constant throughout the domain, which is described by the mass continuity equation:
\begin{equation} \label{eq:incompressible:flow}
    \nabla \cdot \mathbf{u} = 0, \quad \forall \mathbf{x} \in \Omega_t,\forall t\in(0,T).
\end{equation}
The evolution of the free surface \( \eta \) is governed by
the free surface kinematic boundary condition given by:
\begin{equation} \label{eq:free:surface}
    \frac{\partial \eta}{\partial t} = w - u\frac{\partial \eta}{\partial x} - v \frac{\partial \eta}{\partial y}.
\end{equation}
To obtain an equation for the pressure, we assume that \( \mathbf{u} \) is smooth enough so that the derivatives commute. Taking the divergence of both sides of \eqref{eq:incompressible:flow} yields the geometric conservation law:
\begin{equation}
    \rho\frac{\partial}{\partial t}\nabla\cdot\mathbf{u}
    =\mu\nabla\cdot(\nabla^2\mathbf{u}) +\nabla \cdot \mathbf{F} - \nabla^2 p - \rho\nabla\cdot(\mathbf{u}\cdot\nabla)\mathbf{u}.
\end{equation}
If \( \nabla\cdot\mathbf{u}=0 \) for \( t=0 \) and \( \frac{\partial }{\partial t}\nabla\cdot\mathbf{u}=0 \) for all \( t\in(0,T) \), then \( \nabla\cdot\mathbf{u}=0 \) for all \( t\in(0,T) \), and hence the flow is incompressible. Rearranging the above and using \( \nabla\cdot\mathbf{u}=0 \), we obtain the Poisson problem  linking velocity and pressure:
\begin{equation} \label{eq:pressure}
    \nabla^2 p = \mu \nabla\cdot (\nabla^2 \mathbf{u})+\nabla\cdot\mathbf{F} - \rho \nabla\cdot (\mathbf{u} \cdot \nabla) \mathbf{u}.
\end{equation}
We assume \( p = 0 \) on the free surface, which corresponds to a homogeneous Dirichlet boundary condition. We add that it is possible to introduce a decomposition of the pressure into static and dynamic components such that $p=p_s+p_d$, which can be used to define the pressure problem in terms of the dynamic pressure $p_d$; see, for example, \cite{Li2001,Melander2025}. Moreover, we assume that the boundary of the fluid domain is impermeable. We can describe this mathematically by \( \mathbf{n}\cdot\mathbf{u}=0 \), where \( \mathbf{n} \) is the unit of the outward pointing normal vector of the boundary \( \partial \Omega \). By projecting \eqref{eq:navier:stokes} onto \( \mathbf{n} \) and rearranging, we obtain the following condition:
\begin{equation} \label{eq:pressure:neumann}
    \mathbf{n}\cdot\nabla p
    =\mu\mathbf{n}\cdot\nabla^2\mathbf{u} + \mathbf{n}\cdot\mathbf{F} - \rho\mathbf{n}\cdot(\mathbf{u}\cdot\nabla)\mathbf{u},
\end{equation}
where \( \mathbf{n}\cdot\frac{\partial\mathbf{u}}{\partial t}=0 \) because \( \mathbf{n}\cdot\mathbf{u}=0 \), and it is assumed that the seabed is defined in terms of the still-water depth $h(x,y)$, which is assumed to be constant over time.

These equations collectively describe the evolution of non-linear water waves, capturing both the internal fluid dynamics and the coupled interaction with the temporally evolving free surface.

\subsection{\( \sigma \)-Coordinate Transformation}
\label{sec:sigma:transform}

To handle the time dependence of the spatial domain, we introduce a coordinate transformation known as the \(\sigma\)-transform, which maps the evolving physical domain onto a fixed reference domain. The reference coordinates \(((\chi, \gamma, \sigma), \tau) \in \mathbb{R}^3 \times (0,T)\) are defined as
\begin{equation} \label{eq:sigma:transform}
    \chi = x, \quad \gamma = y, \quad \sigma = \frac{z + h(x,y)}{d(x,y,t)}, \quad \tau = t,
\end{equation}
where \( d(x,y,t) = h(x,y) + \eta(x,y,t) \) is the total depth of the water column at time \( t \). This transformation modifies only the vertical coordinate \( z \), mapping it to the fixed reference coordinate \( \sigma \), which satisfies \( 0 \leq \sigma \leq 1 \), with \( \sigma = 1 \) on the free surface and \( \sigma = 0 \) at the bottom. As the reference domain remains fixed in time, we simply denote it by \( \Omega \).

\newpage
The \(\sigma\)-transformation is bijective and is assumed to be sufficiently smooth that the chain rule may be applied to rewrite the governing equations in the transformed coordinates. Using the chain rule, we obtain the "push-forward" of the
differential operators appearing in \eqref{eq:navier:stokes}-\eqref{eq:pressure:neumann}:
\begin{subequations}
\begin{align}
    \frac{\partial}{\partial t} &= \frac{\partial}{\partial \tau} + \frac{\partial \sigma}{\partial t} \frac{\partial}{\partial \sigma}, \\
    \nabla &= \left( \frac{\partial}{\partial \chi} + \frac{\partial \sigma}{\partial x} \frac{\partial}{\partial \sigma}, \quad \frac{\partial}{\partial \gamma} + \frac{\partial \sigma}{\partial y} \frac{\partial}{\partial \sigma}, \quad \frac{\partial \sigma}{\partial z} \frac{\partial}{\partial \sigma} \right).
\end{align}
\end{subequations}
To indicate that the gradient operator acts in the transformed coordinate system, we adopt the notation \( \nabla_\sigma \) for the "push-forward" derived above; however, this is purely notational. With this, the momentum equation \eqref{eq:navier:stokes} in the \(\sigma\)-coordinates becomes
\begin{equation} \label{eq:navier:stokes:sigma}
    \rho \frac{\partial \mathbf{u}}{\partial \tau} + \rho \frac{\partial \sigma}{\partial t} \frac{\partial \mathbf{u}}{\partial \sigma} = -\nabla_\sigma p + \mu \nabla_\sigma^2 \mathbf{u} + \mathbf{F} - \rho \left( \mathbf{u} \cdot \nabla_\sigma \right) \mathbf{u}, \quad \forall (\chi,\gamma,\sigma) \in \Omega, \quad \forall \tau \in (0,T).
\end{equation}
Similarly, the Poisson problem for pressure \eqref{eq:pressure} in the \( \sigma \)-transformed coordinates becomes
\begin{equation*}
    \nabla_\sigma^2 p = \mu \nabla_\sigma \cdot (\nabla_\sigma^2 \mathbf{u}) + \nabla_\sigma \cdot \mathbf{F} - \rho \nabla_\sigma \cdot \left( (\mathbf{u} \cdot \nabla_\sigma) \mathbf{u} \right), \quad \forall (\chi,\gamma,\sigma).
\end{equation*}

At the free surface, the pressure is assumed to be atmospheric, resulting in the homogeneous Dirichlet boundary condition \( p = 0 \) at \( \sigma = 1 \). At solid boundaries, a Neumann condition is obtained by projecting \eqref{eq:navier:stokes:sigma} onto the outward unit normal \( \mathbf{n} \) at a boundary segment \( \Gamma_W \subset \partial \Omega \),
\begin{equation} \label{eq:poisson:neumann:sigma}
    \mathbf{n}\cdot \nabla_\sigma p = \mu \nu \cdot \nabla_\sigma^2 \mathbf{u} + \mathbf{n} \cdot \mathbf{F} - \rho \mathbf{n} \cdot (\mathbf{u} \cdot \nabla_\sigma) \mathbf{u}, \quad x \in \Gamma_W.
\end{equation}

Using the chain-rule and the fact that \( \eta \) does not depend on \( z \), the kinematic boundary condition for the free surface \eqref{eq:free:surface} in the transformed coordinates becomes:
\begin{equation} \label{eq:free:surface:sigma}
    \frac{\partial\eta}{\partial\tau}=w-u\frac{\partial\eta}{\partial\chi}-v\frac{\partial\eta}{\partial\gamma}, \quad z=\eta.
\end{equation}
Since \( \sigma \) is not known a priori, computing its derivatives with respect to \( x, y, z, t \) is not trivial. However, we can take advantage of the fact that both \( \eta \) and \( h \) are independent of the vertical coordinate \( z \), which simplifies the computation. Specifically, we can express the derivatives of \( \sigma \) with respect to the physical coordinates in terms of known quantities by applying the chain and quotient rules. This gives the following expressions:
\begin{subequations}
\begin{align}
    \frac{\partial \sigma}{\partial t} &= -\frac{z + h}{d^2} \frac{\partial d}{\partial t} = -\frac{\sigma}{d} \frac{\partial \eta}{\partial t} = -\frac{\sigma}{d} \frac{\partial \eta}{\partial \tau}, \\
    \frac{\partial \sigma}{\partial x} &= \frac{1}{d} \frac{\partial h}{\partial x} - \frac{z + h}{d^2} \frac{\partial d}{\partial x} = \frac{1}{d} \left( \frac{\partial h}{\partial \chi} - \sigma \frac{\partial d}{\partial \chi} \right), \\
    \frac{\partial \sigma}{\partial y} &= \frac{1}{d} \frac{\partial h}{\partial y} - \frac{z + h}{d^2} \frac{\partial d}{\partial y} = \frac{1}{d} \left( \frac{\partial h}{\partial \gamma} - \sigma \frac{\partial d}{\partial \gamma} \right), \\
    \frac{\partial \sigma}{\partial z} &= \frac{1}{d}.
\end{align}
\end{subequations}

\section{Numerical Discretization} \label{sec:discretization}

Let the tuple \( q = (\mathbf{u}, p, \eta) \) represent the state at time \( \tau \), where \( \mathbf{u} \) is the velocity field, \( p \) the pressure, and \( \eta \) the free surface elevation. The components of \( \mathbf{u} \) and \( p \) are discretized as point values on a structured grid with \( m \) nodes, while \( \eta \) is represented on the subgrid at the free surface corresponding to \( \sigma = 1 \). The subgrid forms a structured grid with codimension one.

We discretize the governing equations \eqref{eq:navier:stokes}-\eqref{eq:pressure:neumann} using the method-of-lines approach proposed by Melander et al.~\cite{Melander2025}. Spatial derivatives are approximated using high-order finite-difference schemes, and the resulting semi-discrete systems for \( \mathbf{u} \) and \( \eta \) are integrated in time using an \( s \)-stage low-storage explicit Runge-Kutta method.

\newpage
For the \( i \)'th stage at time \( \tau_n \), let \( q_n^{(i)} = (\mathbf{u}_n^{(i)}, p_n^{(i)}, \eta_n^{(i)}) \) denote the current state. The right-hand sides of the semi-discrete equations at stage \( i \), corresponding to \eqref{eq:navier:stokes:sigma} and \eqref{eq:free:surface:sigma}, are given by:

\begin{subequations}
\begin{align}
    \mathbf{f}_{\mathbf{u}}(\mathbf{u}_n^{(i)}, p_n^{(i)}, \eta_n^{(i)}, \tau_n^{(i)}) &= -\frac{1}{\rho} \nabla_\sigma^{(i)} p_n^{(i)} + \frac{\mu}{\rho} (\nabla_\sigma^{(i)})^2 \mathbf{u}_n^{(i)} + \frac{1}{\rho} \mathbf{F} - (\mathbf{u}_n^{(i)} \cdot \nabla_\sigma^{(i)}) \mathbf{u}_n^{(i)} + \frac{\sigma}{d} (D_\tau \eta_n^{(i)}) (D_\sigma^{(i)} \mathbf{u}_n^{(i)}), \\
    f_\eta(\mathbf{u}_n^{(i)}, \eta_n^{(i)}, \tau_n^{(i)}) &= w_n^{(i)} - u_n^{(i)} D_\chi \eta_n^{(i)} - v_n^{(i)} D_\gamma \eta_n^{(i)}.
\end{align}
\end{subequations}

where \( \nabla_\sigma^{(i)} \) and \( D_\sigma^{(i)} \) are the \( \sigma \)-transformed finite-difference operators acting on the full grid corresponding to the surface elevation \( \eta_n^{(i)} \), while \( D_\chi \) and \( D_\gamma \) are fixed operators acting on the free surface subgrid. As discussed in Section~\ref{sec:sigma:transform}, it is not necessary to transform the differential operators at the free surface variables in the $xy$-plane.

A general low-storage explicit Runge-Kutta method~\cite{williamson1980} produces updates according to

\begin{subequations}
\begin{align}
    \mathbf{u}_n^{(i)} &= \mathbf{u}_n^{(i-1)} + \beta_k \mathbf{K}_{\mathbf{u}}^{(i)}, &
    \mathbf{K}_{\mathbf{u}}^{(i)} &=
    \begin{cases}
        \mathbf{0}, & i = 0, \\
        \alpha_i \mathbf{K}_{\mathbf{u}}^{(i-1)} + \Delta \tau \, \mathbf{f}_{\mathbf{u}}(\mathbf{u}_n^{(i-1)}, p_n^{(i-1)}, \eta_n^{(i-1)}, \tau_n + c_i \Delta \tau), & i \geq 1,
    \end{cases} \\
    \eta_n^{(i)} &= \eta_n^{(i-1)} + \beta_i K_\eta^{(i)}, &
    K_\eta^{(i)} &=
    \begin{cases}
        0, & i = 0, \\
        \alpha_i K_\eta^{(i-1)} + \Delta \tau \, f_\eta(\mathbf{u}_n^{(i-1)}, \eta_n^{(i-1)}, \tau_n + c_i \Delta \tau), & i \geq 1.
    \end{cases} \\
    \tau_n^{(i)} &= \tau_n + c_i\Delta\tau,
\end{align}
\end{subequations}

where the coefficients \( \alpha_i \), \( \beta_i \), and \( c_i \) depend on the chosen Runge-Kutta scheme, and \( \Delta \tau \) is the time step.  To ensure mass conservation, a discrete Poisson equation for pressure is used to establish a proper velocity-pressure coupling at every stage, ensuring that the discrete velocity field remains divergence-free:

\begin{equation*}
    \nabla_\sigma^{(i)} \cdot \mathbf{u}_n^{(i)} = 0.
\end{equation*}

We obtain a condition for the pressure \( p_n^{(i-1)} \) by substituting the update formula for \( \mathbf{u}_k^{(i)} \) into the divergence free condition:

\begin{align*}
0
&= \nabla_\sigma^{(i)} \cdot \mathbf{u}_n^{(i-1)} + \beta_i \alpha_i \nabla_\sigma^{(i)} \cdot \mathbf{K}_{\mathbf{u}}^{(i-1)} + \beta_k \Delta \tau \nabla_\sigma^{(i)} \cdot \mathbf{f}_{\mathbf{u}}(\mathbf{u}_n^{(i-1)}, p_n^{(i-1)}, \eta_n^{(i-1)}, \tau_n + c_i \Delta \tau) \\
&= \nabla_\sigma^{(i)} \cdot \mathbf{u}_n^{(i-1)}
  + \beta_i \alpha_i \nabla_\sigma^{(i)} \cdot \mathbf{K}_{\mathbf{u}}^{(i-1)} \\
&\quad + \Delta\tau \beta_i \nabla_\sigma^{(i)} \cdot \Bigl(
    -\frac{1}{\rho} \nabla_\sigma^{(i-1)} p_n^{(i-1)}
    + \frac{\mu}{\rho}(\nabla_\sigma^{(i-1)})^2 \mathbf{u}_n^{(i-1)} \\
&\qquad\qquad\qquad
    + \frac{1}{\rho} \mathbf{F}
    - (\mathbf{u}_n^{(i-1)} \cdot \nabla_\sigma^{(i-1)}) \mathbf{u}_n^{(i-1)} \\
&\qquad\qquad\qquad
    + \frac{\sigma}{d}(D_\tau \eta_n^{(i-1)}) (D_\sigma^{(i-1)} \mathbf{u}_n^{(i-1)})
  \Bigr).
\end{align*}

The expression above involves the operator \( \nabla_\sigma^{(i)}\cdot\nabla_\sigma^{(i-1)} \), and isolating for the pressure yields the \textit{mixed-stage} Poisson problem:

\begin{equation} \label{eq:discrete:pressure}
\begin{aligned}
\nabla_\sigma^{(i)} \cdot \nabla_\sigma^{(i-1)} p_n^{(i-1)}
&= \frac{\rho}{\beta_i \Delta \tau} \nabla_\sigma^{(i)} \cdot \mathbf{u}_n^{(i-1)}
 + \frac{\rho \alpha_i}{\Delta \tau} \nabla_\sigma^{(i)} \cdot \mathbf{K}_{\mathbf{u}}^{(i-1)} \\
&\quad + \nabla_\sigma^{(i)} \cdot \Bigl(
    \mu (\nabla_\sigma^{(i-1)})^2 \mathbf{u}_n^{(i-1)}
    + \mathbf{F} \\
&\qquad\qquad
    - \rho (\mathbf{u}_n^{(i-1)} \cdot \nabla_\sigma^{(i-1)}) \mathbf{u}_n^{(i-1)} \\
&\qquad\qquad
    + \rho \frac{\sigma}{d}
      (D_\tau \eta_n^{(i-1)})
      (D_\sigma^{(i-1)} \mathbf{u}_n^{(i-1)})
  \Bigr).
\end{aligned}
\end{equation}

As discussed in section \ref{sec:navier:stokes}, we impose a homogeneous Dirichlet condition \( p = 0 \) at the free surface \( \sigma = 1 \) and a Neumann boundary condition at the solid boundaries \( \Gamma_W \subset \partial \Omega \) by projecting the velocity update onto the normal direction \( \mathbf{n} \). Isolating for the pressure yields:

\begin{equation} \label{eq:discrete:neumann}
\begin{aligned}
\mathbf{n} \cdot \nabla_\sigma^{(i-1)} p_n^{(i-1)}
&= \frac{\rho}{\beta_i \Delta \tau}\, \mathbf{n} \cdot \mathbf{u}_n^{(i-1)}
 + \frac{\rho \alpha_i}{\Delta \tau}\, \mathbf{n} \cdot \mathbf{K}_{\mathbf{u}}^{(i-1)} \\
&\quad + \mathbf{n} \cdot \Bigl(
    \mu (\nabla_\sigma^{(i-1)})^2 \mathbf{u}_n^{(i-1)}
    + \mathbf{F} \\
&\qquad\qquad
    - \rho (\mathbf{u}_n^{(i-1)} \cdot \nabla_\sigma^{(i-1)}) 
      \mathbf{u}_n^{(i-1)} \\
&\qquad\qquad
    + \rho \frac{\sigma}{d}
      (D_\tau \eta_n^{(i-1)})
      (D_\sigma^{(i-1)} \mathbf{u}_n^{(i-1)})
  \Bigr).
\end{aligned}
\end{equation}

The Neumann boundary condition is enforced through ghost nodes, such that these are satisfied together with the governing equations at all grid points simultaneously, as proposed by Engsig-Karup et al. \cite{engsig-karup2009}. The mixed-stage Poisson problem \eqref{eq:discrete:pressure} with the homogeneous Dirichlet boundary condition and the Neumann boundary condition in \eqref{eq:discrete:neumann} defines the system of equations:

\begin{equation} \label{eq:mixed:stage:poisson}
    A_n^{(i)}p_n^{(i)}=b_n^{(i)}.
\end{equation}

To set up the mixed-stage Poisson problem for \( p_n^{(i-1)} \), it is sufficient to know \( \mathbf{u}_n^{(i-1)} \) and \( \eta_n^{(i-1)} \). From \( \mathbf{u}_n^{(i-1)} \) and \( \eta_n^{(i-1)} \), we can compute \( \eta_n^{(i)} \), which is necessary to compute the \(\sigma\)-transformed operator \( \nabla_\sigma^{(i)} \). The remaining terms appearing in \eqref{eq:discrete:pressure} and \eqref{eq:discrete:neumann} have been used to calculate \( \mathbf{u}_n^{(i-1)} \). The derived scheme is described by Algorithm \ref{alg:time:integration}.

\begin{algorithm}
    \caption{Temporal Integration using a Runge-Kutta Method for Incompresible Navier-Stokes with a Free Surface}
    \label{alg:time:integration}
    \begin{algorithmic}[1]
        \Require \( q_n^{(0)}=(\mathbf{u}_n^{(0)},p_n^{(0)},\eta_n^{(0)}) \)

        \State \( \mathbf{K}_\mathbf{u}^{(0)}=0 \)
        \State \( K_\eta^{(0)}=0 \)
        \For{ \( i \in \{1,2,\dots,s\}\) }
            \State \( K_\eta^{(i)}=\alpha_k K_\eta^{(i-1)}+\Delta\tau f_\eta(\mathbf{u}_n^{(i-1)},\eta^{(i-1)},\tau_n+c_i\Delta\tau) \)
            \State \( \eta_n^{(i)} = \eta^{(i-1)}+\beta_i K_\eta^{(i)} \)

            \State \( p_n^{(i-1)} = (A_n^{(i-1)})^{-1}b_n^{(i-1)} \)
                        
            \State \( \mathbf{K}_\mathbf{u}^{(i)} = \alpha_i \mathbf{K}_\mathbf{u}^{(i-1)}+\Delta\tau\mathbf{f}_\mathbf{u}(\mathbf{u}_n^{(i-1)},p_n^{(i-1)},\eta^{(i-1)},\tau_n+c_i\Delta\tau) \)
            \State \( \mathbf{u}^{(i)}=\mathbf{u}^{(i-1)}+\beta_i\mathbf{K}_\mathbf{u}^{(i)} \)
        \EndFor
    \end{algorithmic}
\end{algorithm}

Solving the mixed-stage Poisson problem in \eqref{eq:mixed:stage:poisson} represents the main computational bottleneck in the numerical scheme.
With this discretization, the coordinate-transformed Poisson \eqref{eq:pressure}-\eqref{eq:pressure:neumann} can be expressed as a
time-dependent linear system of equations:
\begin{equation} \label{eq:linear:system}
    A(\tau_n^{(i)})\mathbf{x}(\tau_n^{(i)})=\mathbf{b}(\tau_n^{(i)}).
\end{equation}
The matrix \( A(\tau_n^{(i)}) \) is sparse as a result of the finite-difference discretization in space. Moreover, because of the \( \sigma \)-coordinate transformation \( A(\tau_n^{(i)}) \), it is non-symmetric. Hence, the generalized minimum residual (GMRES) method due to Saad et al.~\cite{saad1986} is a suitable choice to solve the linear systems of equations \eqref{eq:linear:system}. Motivated by the result of Engsig-Karup \cite{engsig-karup2014}, we left-precondition the system with a fixed incomplete \( LU \)-factorization of \( A \) defined in terms of an almost still-water state with waves of very small amplitude (\( \eta/L\ll \mathcal{O}(1) \)) corresponding to a setting \( \eta=0 \) when generating the linear system for such a state.

\newpage
\section{Subspace Acceleration} \label{sec:subspace:acceleration}

To solve the time-dependent system of linear equations in \eqref{eq:linear:system} consecutively using an appropriate iterative solver strategy, we propose using subspace acceleration techniques due to Guido et al.\cite{guido2024} to refine the initial guess. Specifically, we construct a history matrix of previous solutions:
\begin{equation*}
    X(\mathcal{I}) = [\mathbf{x}(\tau_n^{(i)})]_{(n,i)\in\mathcal{I}} \in \mathbb{R}^{m \times K},
\end{equation*}
by concatenating previous solutions from the Runge-Kutta stages at times \( \tau_n^{(i)} \) as columns of \( X(\mathcal{I}) \), ranging over a finite index set \( \mathcal{I} \) with cardinality \( |\mathcal{I}|=K \). Since the history may contain redundant information, we extract a lower-dimensional subspace \( \mathcal{S} \subseteq \mathrm{span}(X(\mathcal{I})) \). The initial guess \( \mathbf{x}_0 \) is then computed by solving the following linear least-squares problem:
\begin{equation*}
    \mathbf{x}_0 = \arg\min_{\mathbf{s} \in \mathcal{S}} \|A(\tau_n^{(i)})\mathbf{s} - \mathbf{b}(\tau_n^{(i)})\|_2^2,
\end{equation*}
where \( \|\cdot\|_2 \) denotes the Euclidean norm. These ideas are summarized in Algorithm \ref{alg:linear:system}, which provides a general template for the strategy. The specific choices required at each step will be detailed in this section.

\begin{algorithm}
    \caption{Solution of linear system of equations, cf. eq. \eqref{eq:linear:system}}
    \label{alg:linear:system}
    \begin{algorithmic}[1]
        \Require \( X(\mathcal{I}) = [\mathbf{x}(\tau_{n}^{(i)})]_{(n,i)\in\mathcal{I}}\in\mathbb{R}^{m\times K} \), \( k\in\mathbb{N}\cap[1,K] \)
        \State Generate a basis \( V\in\mathbb{R}^{m\times k} \) for \( \mathcal{S}\subseteq\mathrm{span}{(X(\mathcal{I}))} \)
        \State Solve \( \mathbf{z}^\star = \arg \min_{\mathbf{z} \in \mathbb{R}^k} \|A(\tau_n^{(i)})V\mathbf{z} - \mathbf{b}(\tau_n^{(i)})\|_2^2 \) 
        \State Solve \( A(\tau_n^{(i)})\mathbf{x}(\tau_n^{(i)}) = \mathbf{b}(\tau_n^{(i)}) \) iteratively with initial guess \( \mathbf{x}_0= V\mathbf{z}^\star \)
    \end{algorithmic}
\end{algorithm}

Algorithm~\ref{alg:linear:system} extends the technique proposed by Guido et al.~\cite{guido2024} by relaxing the requirement that \( V \) be an orthonormal basis. This relaxation significantly accelerates the construction of the subspace, as demonstrated in Section~\ref{sec:numerical:results}. Moreover, it generalizes the projection-based method introduced by Fischer~\cite{Fischer1998}. To see this, assume that \( A \) is time-independent, let \( k = K \), and set \( V = X(\mathcal{I}) \). Since \( A \) is time-independent, it follows that \( AX(\mathcal{I}) = B(\mathcal{I}) \), where \( B(\mathcal{I}) = [\mathbf{b}(\tau_n^{(i)})]_{(n,i) \in \mathcal{I}} \). Hence,
\begin{equation*}
    \mathbf{z}^\star
    = \arg\min_{\mathbf{z} \in \mathbb{R}^k} \| AV\mathbf{z} - \mathbf{b}(\tau_n^{(i)}) \|_2^2
    = \arg\min_{\mathbf{z} \in \mathbb{R}^k} \| B(\mathcal{I})\mathbf{z} - \mathbf{b}(\tau_n^{(i)}) \|_2^2.
\end{equation*}
It follows that \( B(\mathcal{I})\mathbf{z}^\star \) is the orthogonal projection of \( \mathbf{b}(\tau_n^{(i)}) \) onto the range of \( B(\mathcal{I}) \), and the corresponding vector \( \mathbf{x}_0 = V\mathbf{z}^\star \) satisfies \( A\mathbf{x}_0 = B(\mathcal{I})\mathbf{z}^\star \). This coincides with the initial guess construction proposed by Fischer~\cite{Fischer1998}.

The compression of \( X(\mathcal{I}) \) in Step 1 of Algorithm~\ref{alg:linear:system} serves two purposes. First, it reduces the computational cost of solving the linear least-squares problem from \( \mathcal{O}(mK^2) \) to \( \mathcal{O}(mk^2) \) when using a \( QR \)-factorization. Second, it eliminates redundant information in \( X(\mathcal{I}) \), addressing an issue noted by Austin et al.~\cite{austin2021}: the solutions may be close to one another, leading to a highly ill-conditioned matrix. Compressing \( X(\mathcal{I}) \), for example, via singular value decomposition, mitigates this by reducing its condition number.

We also experimented with approximating the solution to the linear least squares problem in Step 2 of Algorithm \ref{alg:linear:system} using the sketch-and-solve approach. However, for the problems considered in this paper, sketching the linear system proved to be more computationally expensive than directly solving the least-squares problem using an \( QR \)-factorization. For other applications and more optimized implementations, sketch-and-solve may still lead to improvements, especially if \( A(\tau_n^{(i)}) \) has some structure that can be exploited by the sketching matrix.

\subsection{Subspace Generation}

A possible strategy to construct a lower-dimensional subspace of \( \mathrm{span}{(X(\mathcal{I})}) \) is to compute an orthonormal basis \( Q\in\mathbb{R}^{m\times k} \) for \( \mathcal{S} \) using the left singular vectors from the truncated singular value decomposition (SVD) of \( X(\mathcal{I}) \). It is well known from the Eckart-Young-Mirsky Theorem that this choice is optimal in the sense that
\begin{equation*}
    \|(I-\Pi)X(\mathcal{I})\|=\min_{\mathrm{rank}(Z)\leq k}\|X(\mathcal{I})-Z\|,
\end{equation*}
where \( \Pi=QQ^\top \) is an orthogonal projection, and \( \|\cdot\| \) is any unitarily invariant norm; for example, the Frobenius or spectral norm. However, computing the truncated SVD is expensive, as explained by Guido et al. \cite{guido2024} and significantly more than we need (both the singular values and the right singular vectors are not needed). Instead, we propose using techniques from randomized linear algebra to construct the subspace and compare it with the optimal subspace constructed from the truncated SVD.

\subsubsection{Randomized Range Finder}

The strategy proposed by Guido et al. \cite{guido2024} uses the \( K \) previous solutions from the Runge-Kutta stages. 
A randomized range finder approach outlined in Algorithm \ref{alg:range:finder} is used to construct the orthonormal basis \( V  = Q\) for the subspace spanned by \( X(\mathcal{I})\Omega \), where \( \Omega\in\mathbb{R}^{n\times k} \) is a sketching matrix for \( k<K\). Crucially, sketching the matrix reduces the dimension of the subspace, which reduces the complexity of orthonormalizing \( X\Omega \) and solving the linear least squares problem.

\begin{algorithm}[h!]
    \caption{Randomized Range Finder}
    \label{alg:range:finder}
    \begin{algorithmic}[1]
        \Require \( X = X(\mathcal{I})\in\mathbb{R}^{m\times K} \), \( k\in\mathbb{N}\cap[1,K] \)
        \State Draw \( \Omega \in \mathbb{R}^{K \times k} \) Gaussian matrix
        \State Sketch \( Y = X \Omega \)
        \State Orthogonalize \( QR = Y \) via \( QR \)-decomposition (optional)
    \end{algorithmic}
\end{algorithm}

The randomized range finder aims to capture most of the range of \( X \) by sampling its column-space. It is a classical result from randomized linear algebra (see Halko et al.\cite{halko2011} Theorem 10.6) that the orthogonal projection \( \Pi_{X\Omega}=QQ^\top \) onto the range of \( X\Omega \) satisfies.
\begin{equation} \label{eq:range:finder}
    \mathbb{E}\|(I-\Pi_{X\Omega})X\|_2\leq \left(1+\sqrt{\frac{p}{q-1}}\right)\min_{\mathrm{rank}(Z)\leq
    k}\|X-Z\|_2+\frac{e\sqrt{p+q}}{q}\min_{\mathrm{rank}(Z)\leq k}\|X-Z\|_F, 
\end{equation}
where we partition \( k \) into \( k = p + q\) for \( p,q\in\mathbb{N} \) and \( q \geq 2 \). We use \( \|\cdot\|_2 \) for the spectral norm and \( \|\cdot\|_F \) for the Frobenius norm. This result shows that, in expectation, the projection \( \Pi_{X\Omega} \) is not much worse than the optimal, provided that the singular values decay rapidly.

In practice, we have observed that it is not necessary for subspace acceleration to orthonormalize the sketch \( Y=X\Omega \), and it is sufficient to take \( V=Y \) as \( \mathrm{span}(X\Omega) = \mathrm{span}(Q) \) in exact arithmetic. This significantly reduces the computational cost of constructing the subspace from \( \mathcal{O}(mk^2+mkK) \) to \( \mathcal{O}(mkK) \), as the \( QR \)-factorization does not need to be updated at every iteration, whether computed from scratch or using rank-\( 1 \) updates, as discussed by Guido et al.\cite{guido2024}.
The orthonormalization step is included in algorithm \ref{alg:range:finder} because of its connection to the orthogonal projection in the theoretical bound \eqref{eq:range:finder} and the randomized SVD \cite{halko2011}.

However, as we will demonstrate in Section \ref{sec:numerical:results}, small differences have been observed between using \( V=Y \) and \( V=Q \) as the generating set for \( \mathcal{S} \). These discrepancies do not significantly affect performance and can be attributed to numerical round-off errors. For highly ill-conditioned system matrices \( A(\tau) \), it might be necessary to use a well-conditioned basis \( V=Q \).

\subsubsection{Exponentially Weighted Moving Average (EWMA)}

As mentioned, it is possible to update the sketch \( X\Omega \) and \( QR \)-factorization in Algorithm \ref{alg:range:finder} using rank-\( 1 \) updates. However, this requires storing the previous \( K \) solution and the sketching matrix \( \Omega \). For memory-restricted applications, this might pose a serious issue. Moreover, numerical round-off errors from the rank-\( 1 \) updates accumulate; therefore, it is necessary to recompute \( X\Omega \) and \( QR \) from scratch after a number of iterations. 

In order to address these issues, we propose an alternative approach for constructing the subspace described in Algorithm \ref{alg:ewma}. The idea is to use the entire history and to continuously forget it at a prescribed rate \( r \). As we will see in section \ref{sec:numerical:results}, it is crucial to forget previous solutions to produce a good initial guess.

\begin{algorithm}
    \caption{Exponentially Weighted Moving Average}
    \label{alg:ewma}
    \begin{algorithmic}[1]
        \Require \( \mathbf{x}(\tau_1^{(0)}),\dots,\mathbf{x}(\tau_n^{(i)}) \), \( k\in\mathbb{N} \), \( r\in(0,1] \)
        \State Draw \( \omega_{1,0},\dots,\omega_{n,i} \in \mathbb{R}^{k} \) Gaussian vectors
        \State Sketch \( Y = \sum_{(\ell,j)\in\mathcal{I}} r^{\tau_n^{(i)}-\tau_\ell^{(j)}}\mathbf{x}(\tau_\ell^{(j)})\omega_{\ell,j}^\top \)
        \State Orthonormalize \( QR=Y \) via \( QR \)-decomposition (optional)
    \end{algorithmic}
\end{algorithm}

The main advantage of Algorithm \ref{alg:ewma} is that it is not necessary to store the full history, and the sketch \( Y \) can be updated with rank-\( 1 \) updates without accumulating round-off errors. To see this, let \( Y_n^{(i)} \) denote the sketch at the \( n \)'th time step and \( i \)'th Runge-Kutta stage. We can rewrite \( Y_n \) as follows:
\begin{equation*}
    Y_{n}^{(i)}
    = \sum_{(\ell,j)\in\mathcal{I}} r^{\tau_n^{(i)}-\tau_\ell^{(j)}}\mathbf{x}(\tau_\ell^{(j)})\omega_{\ell,j}^\top
    = r^{\tau_n^{(i)}-\tau_{n'}^{(i')}} Y_{n'}^{(i')}+\mathbf{x}(\tau_n^{(i)})\omega_{n,i}^\top,
\end{equation*}
where \( (n',i') \) is the previous time step and the Runge-Kutta stage. Hence, updating \( Y \) involves \( 2mk \) multiplications and \( mk \) additions, resulting in a total computational complexity of \( \mathcal{O}(mk) \) without orthonormalization. This compares favorably with the computational complexity of Algorithm~\ref{alg:range:finder}, which is at least \( \mathcal{O}(mkK) \). Furthermore, round-off errors in \( Y_n^{(i)} \) are suppressed at a rate \( r \), eliminating the need to recompute \( Y \) from scratch. This also reduces the minimum memory requirements compared to Algorithm~\ref{alg:range:finder}. As with Algorithm~\ref{alg:range:finder}, we find that orthogonalization of \( Y \) is not necessary in practice, and simply setting \( V = Y \) yields satisfactory results.

\subsection{Convergence Results}

Guido et al.\cite{guido2024} provide a detailed analysis of the quality of the initial guess constructed from Algorithm \ref{alg:linear:system}. After a reparameterization, we can assume the parameterized linear systems are of the form  
\begin{equation*}
    A(t)\mathbf{x}(t) = \mathbf{b}(t), \quad A : [-1,1] \to \mathbb{C}^{m \times m}, \quad \mathbf{b} : [-1,1] \to \mathbb{C}^n.
\end{equation*}
Let \( \mathbf{t} = (t_1, \dots, t_K) \) with \( t_1, \dots, t_K \in [-1,1] \) distinct times, and define the history matrix \( X(\mathbf{t}) = [\mathbf{x}(t_1) \, | \, \cdots \, | \, \mathbf{x}(t_K)] \). The analysis exploits the optimality of \( \mathbf{s}^\star=V\mathbf{z}^\star \) in \( \mathcal{S} \) by comparing it with the orthogonal projection of \( \mathbf{x}(t) \) onto \( \mathcal{S} \). However, since there is no guarantee that \( \mathbf{x}(t) \in \mathrm{span}(X(\mathbf{t})) \) holds for all \( t\in[0,1] \), there is no meaningful way to determine how far \( \mathbf{x}(t) \) is from \( \mathcal{S} \). To address this, \( \mathbf{x}(t) \) is approximated at the nodes \( t_1, \dots, t_K \) using a degree-\( d \) vector-valued polynomial \( \mathbf{p}_d \), which takes values in \( \mathrm{span}(X(\mathbf{t})) \) while controlling the extrapolation error.

To control the extrapolation error, it is assumed that \( A \) and \( \mathbf{b} \) satisfy conditions that ensure  \( \mathbf{x}(t) \) admits an analytic extension to the open Bernstein ellipse \( E_\rho \subset \mathbb{C} \) for \( \rho > 1 \), with foci at \( \pm 1 \) and semi-major and semi-minor axes summing to \( \rho \). The parameter \( \rho \) is determined by the analytical properties of \( A \) and \( \mathbf{b} \) and controls how far away from \( [-1,1] \) the extrapolation error can be bounded. To control how far away \( \mathbf{p}_d(t) \) is from \( \mathcal{S} \), it is assumed that the orthogonal projection \( \Pi:\mathbb{R}^m\to\mathcal{S} \) onto \( \mathcal{S} \) satisfies
\begin{equation*}
    \|(I - \Pi)X(\mathbf{t})\|_2 \leq \epsilon,
\end{equation*}
for some \( \epsilon\geq0 \). It is important to notice that \( \Pi \) is independent of how \( \mathcal{S} \) was constructed. In view of \eqref{eq:range:finder}, we expect \( \epsilon \) to be small when \( \mathcal{S}=\mathrm{span}{(X(\mathbf{t})\Omega)} \) and the singular values of \( \mathbf{X}(\mathbf{t}) \) decay rapidly. These observations lead to the straightforward generalization of the main result by Guido et al.\cite{guido2024} stated below.

\begin{theorem} \label{thm:initial:guess}
    Let \( \rho>1 \), \( \epsilon\geq0 \), and \( \mathcal{S}\subseteq\mathrm{span}{(X)} \) be a subspace, and let \( \Pi:\mathbb{R}^m\to\mathcal{S} \) be the orthogonal projection onto \( \mathcal{S} \). If \( A:[-1,1]\to\mathbb{C}^{n\times n} \) and \( b:[-1,1]\to\mathbb{C}^n \) admit analytic extensions to \( E_\rho \) such that \( A(t) \) is invertible for all \( t\in \overline{E}_\rho \) and \( \|(I-\Pi)X(\mathbf{t})\|\leq\epsilon \), then the initial guess \( \mathbf{s}^\star \) constructed from Algorithm \ref{alg:linear:system}  satisfies
    \begin{equation} \label{eq:initial:guess}
        \|A(t)\mathbf{s}^\star-\mathbf{b}(t)\|_2\leq \|A(t)\|_2\left(\kappa_\rho C(\mathbf{t},s,\rho,M)+\epsilon D(\mathbf{t},s,\rho)\right)s^{d+1}, \quad \forall t\in E_\rho\cap\mathbb{R}\setminus(-1,1),d\leq K-1,
    \end{equation}
    where \( s=(|t|+\sqrt{|t|^2-1})/\rho<1 \),\( \kappa_p=\max_{t\in \partial E_p}\|\mathbf{x}(t)\|_2 \), \( C(\mathbf{t},s,\rho,K) \), and \( D(\mathbf{t},s,\rho) \) are constants depending on \( \mathbf{t} \), \( s \), \( \rho \), and \( K \).
\end{theorem}

It is readily seen that if \( \mathcal{S}=\mathrm{span}{(X(\mathbf{t}))} \), then Theorem \ref{thm:initial:guess} holds with \( \epsilon=0 \). A special case of Theorem \ref{thm:initial:guess} with equidistant time-steps is provided by Guido et al. \cite{guido2024}.

\section{Numerical Results} \label{sec:numerical:results}

In this section, we compare the different subspace acceleration methods discussed in section \ref{sec:subspace:acceleration} to the numerical simulation of nonlinear water waves using the solver described in section \ref{sec:discretization}. We start by examining the effectiveness of the subspace acceleration applied to the solution of stream function waves and optimize the subspace parameters \( K \), \( k \), and the forgetfulness rate \( r \). We recall that \( K \) is the number of columns in the history matrix, while \( k < K\) is the dimension of the low-dimensional subspace we use to generate the guess.  With these parameters fixed, we examine the effectiveness of subspace acceleration across a range of wave parameters. Finally, we demonstrate the effectiveness of subspace acceleration for a wave tank setup and compare it with experimental data.

All numerical experiments have been performed on a 2021 16-inch MacBook Pro equipped with an Apple M1 Pro processor, 16 GB of RAM, running macOS Sequoia and MATLAB 2023b. The mixed-stage Poisson problem in the algorithm \ref{alg:time:integration} has been solved using GMRES preconditioned with an incomplete \( LU \)-factorization of the \( \sigma \)-transformed Laplacian at still-water depth \( \eta=1 \). The coefficients in Algorithm \ref{alg:time:integration} come from the low-storage explicit 4th order 5-stage Runge-Kutta method (LSERK45) proposed by Carpenter et al.~\cite{Carpenter1994}.

\subsection{Stream function wave propagation}

To evaluate the performance of subspace acceleration, we initialize a stream function wave at \( t = 0 \) with a known surface velocity \( (u_s, w_s) \) and elevation \( \eta \). The stream function wave provides an analytical solution to the potential flow that is inviscid,  irrotational, and incompressible \cite{Engsig-Karup2013}. Solutions can be obtained using the Fourier method proposed by Fenton et al.\cite{fenton1982}. The stream function solution depends on the non-dimensional wave-number \( kh \) (dispersion) and the relative wave steepness \( (H/L)_{\max} \) (nonlinearity), where the latter represents the maximum steepness before wave breaking, as determined by Battjes \cite{battjes1974}:
\begin{equation*}
    \left(\frac{H}{L}\right)_{\max} = 0.1401 \tanh(0.8863kh).
\end{equation*}
This solution provides a solid foundation for evaluating performance across a broad range of regimes--from small- to finite-amplitude wave propagation--and for varying relative depths, including shallow (\( kh < 0.4 \)), intermediate (\( 0.4 < kh < 2 \)), and deep water (\( kh > 2 \)). It also enables a systematic assessment at different levels of wave steepness. In general, steeper nonlinear waves pose the greatest challenges for numerical solvers.


We begin the numerical analysis by optimizing the subspace parameters \( K \), \( k \), and \( r \) to balance the computational cost of computing the initial guess in Algorithm~\ref{alg:linear:system}against the resulting improvement in the guess and, consequently, the convergence rate of GMRES.  
The experiments are performed in a two-dimensional domain that is periodic in the \( x \)-direction and uniformly discretized with \( N_x = 200 \) and \( N_z = 20 \) nodes along the \( x \)- and \( z \)-axes, respectively, with one additional ghost node just below the sea bed. The resulting linear system in~\eqref{eq:linear:system} therefore has size \( N_x (N_z + 1) = 4200 \).  
Time integration is carried out using the LSERK45 scheme with a step size of \( \Delta \tau = 10^{-4} \). The history matrix is initialized by first applying the baseline algorithm for the \( N_\mathrm{init} = 6 \) time steps, followed by the \( N_\mathrm{step} = 200 \) time steps using the proposed strategy. The reported results include only data from the final \( N_\mathrm{step} \) time steps.  
For comparison, the baseline algorithm uses the GMRES solver with the previous solution as an initial guess.



\begin{center}
\begin{table*}[!h]%
\caption{Case: Stream function waves. The average number of iterations, time and speed up per time-step compared to the baseline using Algorithm \ref{alg:range:finder} for different subspace parameters.}
\label{tab:subspace:dimension}
\begin{tabular*}{\textwidth}{@{\extracolsep\fill}cccccccccc@{}}
\toprule
& &\multicolumn{4}{@{}c}{\( V=X\Omega  \)} & \multicolumn{4}{@{}c}{\( V= Q\)} \\

\cmidrule{3-6}\cmidrule{7-10}
\( K \) & \( k \) & Iterations & Speed Up  & Time\([\mathrm{ms}]\) & Speed Up  & Iteration & Speed Up & Time\( [\mathrm{ms}] \) & Speed Up \\
\midrule
15 &  6 & 6.26 &  2.47 & 2.66 & 1.85 & 6.21 &  2.48 & 2.74 & 1.79 \\
15 &  8 & 2.52 &  6.14 & 1.57 & 3.14 & 2.52 &  6.12 & 1.72 & 2.86 \\
15 & 10 & 1.69 &  9.14 & 1.46 & 3.36 & 1.54 & 10.00 & 1.66 & 2.97 \\
20 & 10 & 1.82 &  8.49 & 1.52 & 3.23 & 1.80 &  8.60 & 1.83 & 2.69 \\
20 & 15 & 1.15 & 13.37 & 1.75 & 2.82 & 1.19 & 13.01 & 2.19 & 2.25 \\
30 &  8 & 3.65 &  4.23 & 2.04 & 2.41 & 3.63 &  4.26 & 2.25 & 2.19 \\
30 & 12 & 1.47 & 10.53 & 1.60 & 3.07 & 1.50 & 10.28 & 1.94 & 2.54 \\
30 & 15 & 1.24 & 12.46 & 1.81 & 2.72 & 1.31 & 11.78 & 2.29 & 2.14 \\
\bottomrule
\end{tabular*}
\begin{tablenotes}
\item The experiment uses dispersion \( kh=1.0 \) and steepness = \( H/L=0.60(H/L)_{\max}= 0.0602 \).
\item The baseline algorithm uses an average of \( 15.4 \) GMRES iterations and \( 4.92\mathrm{ms} \).
\end{tablenotes}
    
\end{table*}
\end{center}

Table \ref{tab:subspace:dimension} compares the average number of GMRES iterations and the average computational time per time step. The results clearly demonstrate that subspace acceleration significantly reduces the number of GMRES iterations required to achieve the desired tolerance compared to the baseline.
We also compared with polynomial extrapolation techniques studied by Austin et al. \cite{austin2021}, but found that the number of GMRES iterations increased for the test cases considered in this study. Therefore, we do not report these numbers.

A more detailed analysis of the data in Table \ref{tab:subspace:dimension} reveals several trends.
For a fixed \( K \), the average number of iterations decreases monotonically as \( k \) increases. This behavior is consistent with our intuition, as a larger subspace \( \mathcal{S} \) provides a better approximation space, leading to a consistent reduction in the residual of the initial guess. A similar trend is observed when both \( K \) and \( k \) increase simultaneously. However, increasing \( K \) alone does not decrease the number of GMRES iterations. This suggests that the most recent solutions contain more predictive information about the next solution, which is expected for continuous functions. Consequently, incorporating a forgetting mechanism to discard older solutions is crucial for efficiency.

\newpage
The relationship between subspace parameters and the average time per iteration is more complicated, as it also accounts for the computational cost of constructing the subspaces, which increases with both \( K \) and \( k \). Our results suggest that \( K=15 \) and \( k=10 \) strike a good balance between history size and subspace dimension. We adopt these parameters for the subsequent numerical experiments. Additionally, we observe differences between the methods, despite the subspaces being theoretically identical in exact arithmetic.
We attribute this observation to numerical round-off errors.

\begin{center}
\begin{table*}[!h]%
\caption{Case: Stream function waves. The average number of iterations per time-step using Algorithm \ref{alg:ewma} for different subspace dimensions and forgetness rate.}
\label{tab:forgetful:iterations}
\begin{tabular*}{\textwidth}{@{\extracolsep\fill}cccccccccc@{}}
\toprule
 & 1.0 & 0.95  & 0.9 & 0.85  & 0.8 & 0.75 & 0.7 & 0.65 & 0.6 \\
\midrule
 6 & 24.88 & 4.88 & 4.74 & 4.27 & 4.14 & 3.88 & 3.78 & 3.53 & 3.23 \\
 8 & 17.94 & 2.40 & 2.23 & 2.17 & 2.13 & 2.03 & 2.00 & 1.89 & 1.77 \\
10 & 16.47 & 1.81 & 1.72 & 1.63 & 1.59 & 1.57 & 1.53 & 1.44 & 1.49 \\
12 & 15.69 & 1.53 & 1.46 & 1.56 & 1.39 & 1.31 & 1.32 & 1.29 & 1.31 \\
15 &  9.78 & 1.26 & 1.23 & 1.22 & 1.18 & 1.20 & 1.15 & 1.19 & 1.22 \\
\bottomrule
\end{tabular*}
\begin{tablenotes}
\item The experiment uses dispersion \( kh=1.0 \) and steepness = \( H/L=0.60(H/L)_{\max}\approx 0.0602 \). \( K=15 \) and \( k=10 \) are used to generate the guess.
\end{tablenotes}
\end{table*}
\end{center}

Table \ref{tab:forgetful:iterations} shows a comparison of the average number of GMRES iterations per time step for different subspace dimensions \( k \) and forgetting rates \( r \) when using Algorithm \ref{alg:ewma} to generate the initial guess. The results indicate that, similar to Algorithm \ref{alg:range:finder}, the average number of GMRES iterations decreases monotonically with increasing \( k \) and tends to decrease as \( r \) decreases. The particularly poor performance observed for \( r = 1.0 \) underscores the necessity of forgetting older solutions, as very old solutions may not contain relevant information and the numerical round-off errors are not being suppressed.

For \( r < 1 \), the average number of GMRES iterations remains largely unaffected by variations in \( r \). However, as \( r \to 0 \), we expect the performance to converge to the baseline, as the method effectively discards historical information. Experimental verification of this claim was infeasible, as the sketch \( Y \) from Algorithm \ref{alg:ewma} becomes highly ill-conditioned for small \( r \), eventually reducing to a rank-one matrix. Based on these findings, we recommend selecting \( 0.7 \leq r < 1 \) to balance stability and efficiency.

The corresponding average time per time-step is shown in Table \ref{tab:forgetful:times}. Our results show that we obtain good results for both \( k=8 \) and \( k=10 \). Because \( k=10 \) also yields good results when using Algorithm \ref{alg:range:finder}, we use \( k=10 \) in both cases to make the comparison more similar. For the forgetfulness rate, we choose \( r=0.70 \) in light of the above discussion. We will use these parameters in the following when we consider the performance of Algorithm \ref{alg:ewma} to construct the subspace.

\begin{center}
\begin{table*}[!h]%
\caption{Case: Stream function waves. The average time in milliseconds per time-step using Algorithm \ref{alg:ewma} for different subspaces dimension and forgetness rates.}
\label{tab:forgetful:times}
\begin{tabular*}{\textwidth}{@{\extracolsep\fill}cccccccccc@{}}
\toprule
 & 1.0 & 0.95  & 0.9 & 0.85  & 0.8 & 0.75 & 0.7 & 0.65 & 0.6 \\
\midrule
 6 & 8.27 & 2.17 & 2.12 & 2.05 & 1.91 & 1.87 & 1.83 & 1.73 & 1.65 \\
 8 & 6.06 & 1.52 & 1.48 & 1.51 & 1.43 & 1.50 & 1.43 & 1.36 & 1.34 \\
10 & 5.81 & 1.50 & 1.54 & 1.53 & 1.44 & 1.45 & 1.45 & 1.41 & 1.42 \\
12 & 5.71 & 1.58 & 1.65 & 1.66 & 1.56 & 1.56 & 1.55 & 1.53 & 1.53 \\
15 & 4.21 & 1.77 & 1.82 & 1.79 & 1.76 & 1.94 & 1.76 & 1.76 & 1.75 \\
\bottomrule
\end{tabular*}
\begin{tablenotes}
\item The experiment uses dispersion \( kh=1.0 \) and steepness = \( H/L=0.60(H/L)_{\max}\approx 0.0602 \). \( K=15 \) and \( k=10 \) are used to generate the guess.
\end{tablenotes}
\end{table*}
\end{center}

With the chosen subspace parameters, we compare the performance of different algorithms for constructing subspaces in Table \ref{tab:iteration:comparison} across various wave parameters. Our results indicate that orthonormalization in Algorithm \ref{alg:range:finder} is not necessary and that the EWMA approach performs better in general. Furthermore, we observe that the average number of GMRES iterations increases with \( kh \) and \( H/L \). For lower values of \( kh \) and \( H/L \), the surface elevation is closer to zero, leading to more effective preconditioning and, consequently, fewer GMRES iterations. Therefore, in the small-amplitude setting, we find that there is less room for improving the initial guess for the iterative solver.

\clearpage

\begin{center}
\begin{table*}[h!]%
\caption{Case: Stream function waves. The average number of iterations per time-step and speed up compared to the baseline for different wave-parameters.}
\label{tab:iteration:comparison}
\begin{tabular*}{\textwidth}{@{\extracolsep\fill}ccccccccc@{}}
\toprule
& & Baseline & \multicolumn{2}{@{}c}{\( X\Omega \)} & \multicolumn{2}{@{}c}{Randomized Range Finder} & \multicolumn{2}{@{}c}{EWMA} \\

\cmidrule{3-3}\cmidrule{4-5}\cmidrule{6-7}\cmidrule{8-9}
\( kh \) & \( \frac{H/L}{(H/L)_{\max}} \) & Average & Average & Speed Up  & Average & Speed Up  & Average & Speed Up \\
\midrule
0.4 & 0.01 &  4.85 & 0.60 &  7.50 & 0.61 &  8.59 & 0.36 & 12.46 \\
2.0 & 0.01 & 14.48 & 1.07 & 13.69 & 1.00 & 13.22 & 0.94 & 16.65 \\
4.0 & 0.01 &  7.38 & 1.35 &  5.32 & 1.37 &  5.43 & 1.28 &  5.84 \\
0.4 & 0.30 &  7.13 & 0.89 &  8.45 & 0.88 &  8.48 & 0.81 &  8.56 \\
2.0 & 0.30 & 16.86 & 1.34 & 11.59 & 1.30 & 13.14 & 1.21 & 13.65 \\
4.0 & 0.30 &  8.30 & 1.54 &  4.53 & 1.51 &  4.81 & 1.38 &  5.22 \\
0.4 & 0.70 & 10.39 & 1.59 &  6.48 & 1.57 &  6.63 & 1.39 &  7.27 \\
2.0 & 0.70 & 21.55 & 2.00 & 10.30 & 1.88 & 10.05 & 1.81 & 12.90 \\
4.0 & 0.70 & 21.09 & 2.08 &  8.20 & 1.98 &  9.57 & 1.98 & 10.03 \\
\bottomrule
\end{tabular*}
\begin{tablenotes}
\item The subspace paramters used are \( K=15 \), \( k=10 \) and forgetfullness \( r=0.70 \).
\end{tablenotes}
\end{table*}
\end{center}

\begin{figure}[!h]
    \centering
    \includegraphics[width=\textwidth]{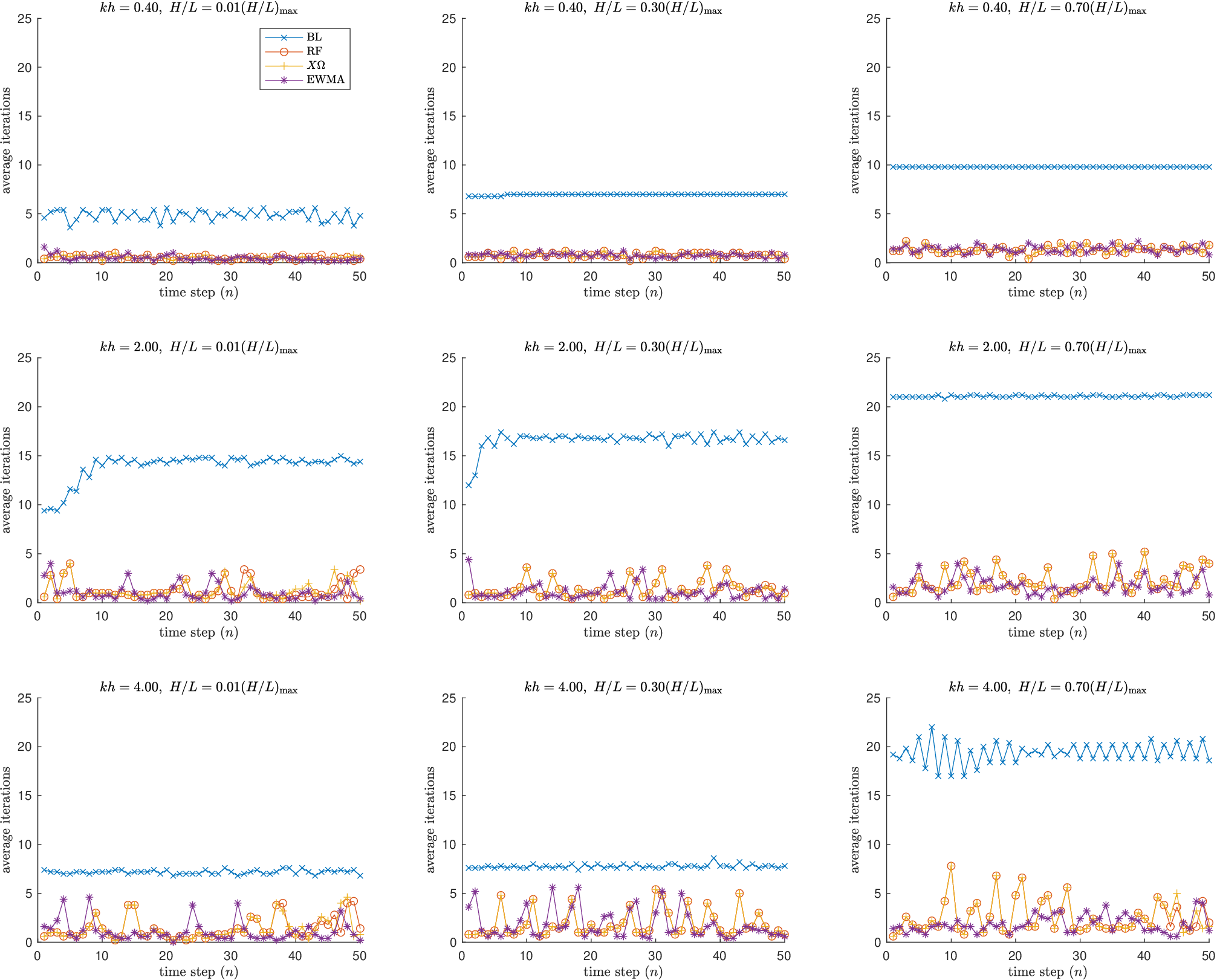}
    \caption{The figure shows the average number of GMRES iterations at every time-step to solve the linear system of equations using an ILU-preconditioned GMRES for different wave parameters.}
    \label{fig:iteration:sfwave}
\end{figure}

\clearpage

\begin{center}
\begin{table*}[h!]%
\caption{Case: Stream function waves. The average time per time-step and speed up compared to the baseline for wave-parameters and subspace parameters.}
\label{tab:time:comparison}
\begin{tabular*}{\textwidth}{@{\extracolsep\fill}ccccccccc@{}}
\toprule
& & Baseline & \multicolumn{2}{@{}c}{\( X\Omega \)} & \multicolumn{2}{@{}c}{Randomized Range Finder} & \multicolumn{2}{@{}c}{EWMA} \\

\cmidrule{3-3}\cmidrule{4-5}\cmidrule{6-7}\cmidrule{8-9}
\( kh \) & \( \frac{H/L}{(H/L)_{\max}} \) & Average \([\mathrm{ms}]\) & Average \([\mathrm{ms}]\) & Speed Up  & Average \([\mathrm{ms}]\) & Speed Up  & Average \([\mathrm{ms}]\) & Speed Up \\
\midrule
0.4 & 0.01 & 1.94 & 1.33 & 1.46 & 1.54 & 1.26 & 1.21 & 1.60 \\
2.0 & 0.01 & 4.55 & 1.42 & 3.22 & 1.58 & 2.88 & 1.35 & 3.37 \\
4.0 & 0.01 & 2.55 & 1.49 & 1.71 & 1.68 & 1.52 & 1.44 & 1.77 \\
0.4 & 0.30 & 2.47 & 1.44 & 1.71 & 1.62 & 1.52 & 1.40 & 1.77 \\
2.0 & 0.30 & 5.23 & 1.49 & 3.51 & 1.66 & 3.15 & 1.40 & 3.72 \\
4.0 & 0.30 & 2.78 & 1.54 & 1.80 & 1.72 & 1.61 & 1.46 & 1.90 \\
0.4 & 0.70 & 3.30 & 1.70 & 1.94 & 1.89 & 1.75 & 1.57 & 2.10 \\
2.0 & 0.70 & 6.68 & 1.66 & 4.02 & 1.81 & 3.69 & 1.57 & 4.25 \\
4.0 & 0.70 & 6.56 & 1.73 & 3.80 & 1.89 & 3.47 & 1.66 & 3.96 \\
\bottomrule
\end{tabular*}
\begin{tablenotes}
\item The subspace paramters used are \( K=15 \), \( k=10 \) and forgetfullness \( r=0.70 \).
\end{tablenotes}
\end{table*}
\end{center}

\begin{figure}[h!]
    \centering
    \includegraphics[width=\textwidth]{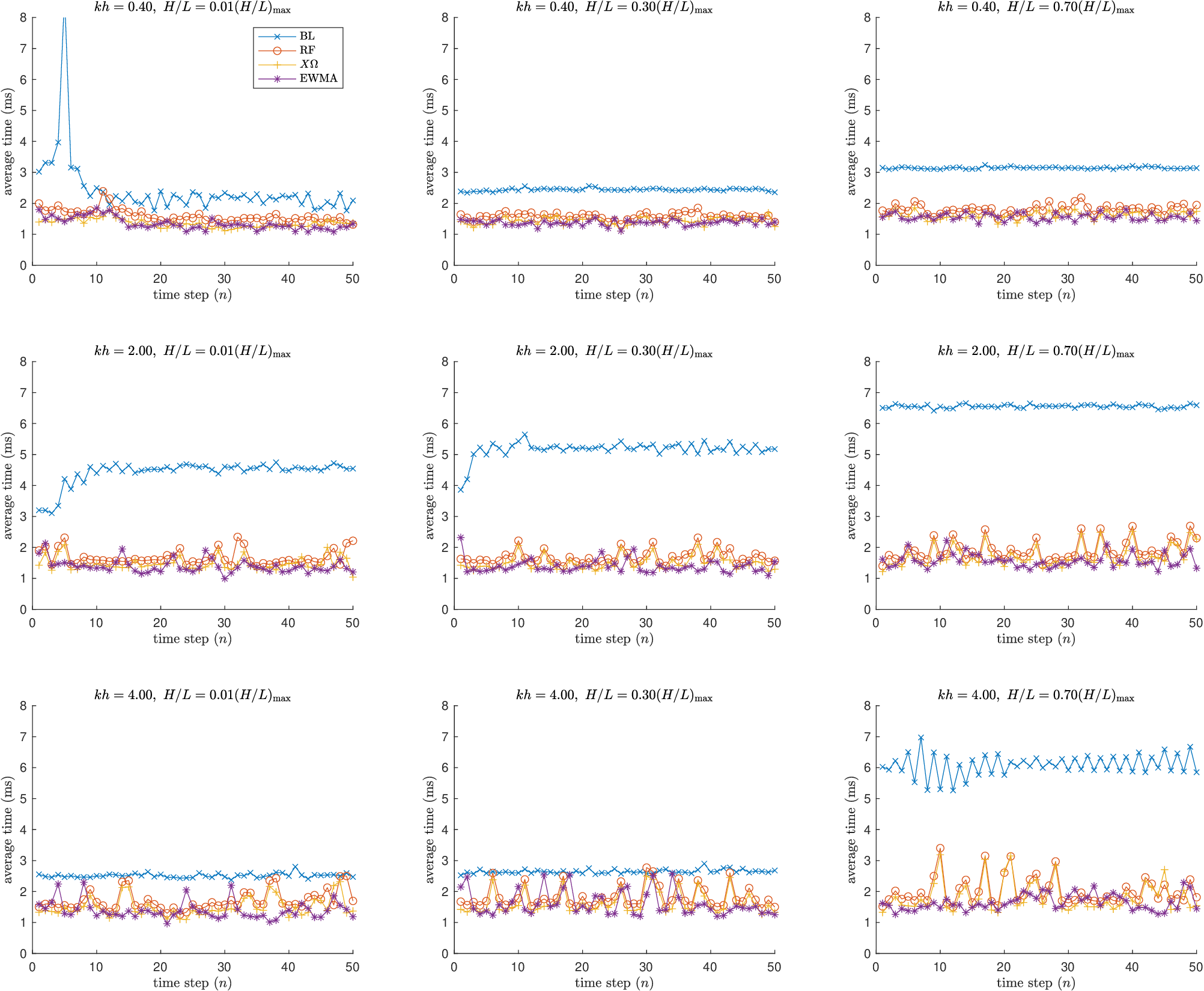}
    \caption{The figure shows the average time used at every time-step to compute the initial guess and to solve the linear system of equations using an ILU-preconditioned GMRES.}
    \label{fig:time:sfwave}
\end{figure}

\clearpage

The most significant reduction in GMRES iterations occurs for \( kh = 2.0 \). This is because the baseline algorithm requires particularly many iterations for \( kh = 2.0 \). This can also be seen in  Figure~\ref{fig:iteration:sfwave}, which shows the number of GMRES iterations at each time step. This suggests that for \( kh = 2.0 \), the current solution is further away from the next solution, whereas for lower and higher values of \( kh \), the opposite is true. Consequently, subspace acceleration appears to be most effective for intermediate depths.
Figure \ref{fig:iteration:sfwave} further illustrates that the initial guesses provided by the subspace acceleration techniques are significantly more accurate than simply using the previous solution. In many cases, these improved initial guesses satisfy the convergence tolerance before GMRES iterations begin, leading to substantial computational savings. It also highlights a significant problem with using LSERK45, which has non-uniform time-steps that lead to the oscillatory behavior of the baseline algorithm seen in Figure \ref{fig:iteration:sfwave}. This pattern is less obvious when using subspace acceleration techniques.

Table \ref{tab:time:comparison} presents the average time per time step required to construct the initial guess and to solve the linear system using GMRES. The numerical results show that the subspace produced by Algorithm \ref{alg:ewma} outperforms Algorithm \ref{alg:range:finder}, both with and without orthonormalization. This outcome is expected, as Algorithm \ref{alg:ewma} efficiently updates the subspace using its streaming property while also producing a better initial guess on average, as indicated by the results in Table \ref{tab:iteration:comparison}.
We also observe that omitting orthonormalization in Algorithm \ref{alg:range:finder} results in a faster overall computation time, despite the fact that the initial guess without orthonormalization leads to a higher average number of GMRES iterations, as seen in Table \ref{tab:iteration:comparison}.  

Figure \ref{fig:time:sfwave} illustrates the computational time in the first \( N_\mathrm{steps} \) iterations, revealing a trend similar to that observed in Figure \ref{fig:iteration:sfwave}. This suggests that GMRES iterations dominate the overall computational cost at each time step, which is consistent with our observations. Furthermore, Algorithm \ref{alg:range:finder} with orthonormalization is consistently better at each time step, although it requires fewer GMRES iterations on average, as seen in Table \ref{tab:iteration:comparison}. Based on these results, it appears to be beneficial to omit the expensive orthonormalization step and simply use the sketch to construct the subspace.

\subsection{Harmonic wave generation over a submerged bar}

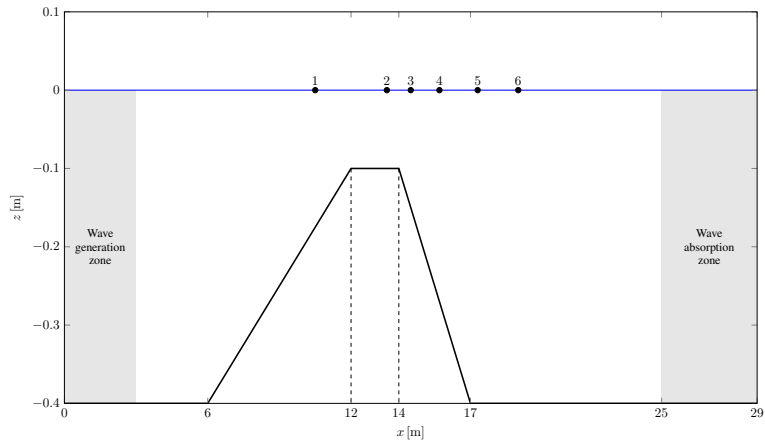
\begin{figure}[!h]
    \centering
    \begin{tikzpicture}[scale=0.5]
    \begin{axis}[
        width=20cm,
        height=12cm,
        xlabel={$x\,[\mathrm{m}]$},
        ylabel={$z\,[\mathrm{m}]$},
        xmin=0, xmax=29,
        ymin=-0.4, ymax=0.1,
        xtick={0,6,12,14,17,25,29},
        ytick={-0.4,-0.3,-0.2,-0.1,0.0,0.1},
        grid style=dashed,
        axis on top=true,
    ]
    
    \fill[gray!20] (axis cs:0,0) rectangle (axis cs:3,-0.4);
    \node[align=center, font=\small] at (axis cs:1.5,-0.2) {Wave\\generation\\zone};
    
    \fill[gray!20] (axis cs:25,0) rectangle (axis cs:29,-0.4);
    \node[align=center, font=\small] at (axis cs:27,-0.2) {Wave\\absorption\\zone};
    
    \addplot[black, very thick] coordinates {
        (-10,-0.4)
        (6,-0.4)
        (12,-0.1)
        (14,-0.1)
        (17,-0.4)
        (29,-0.4)
    };
    
    \addplot[black, dashed] coordinates {
        (12,-0.1)
        (12,-0.4)
    };
    
    \addplot[black, dashed] coordinates {
        (14,-0.1)
        (14,-0.4)
    };
    
    \addplot[blue, thick] coordinates {
        (0,0)
        (29,0)
    };
    
    \addplot[
        only marks,
        mark=*,
        mark size=2pt,
    ] coordinates { (10.5, 0) };
    \node at (axis cs:10.5, 0) [anchor=south] {\( 1 \)};
    
    \addplot[
        only marks,
        mark=*,
        mark size=2pt,
    ] coordinates { (13.5, 0) };
    \node at (axis cs:13.5, 0) [anchor=south] {\( 2 \)};
    
    \addplot[
        only marks,
        mark=*,
        mark size=2pt,
    ] coordinates { (14.5, 0) };
    \node at (axis cs:14.5, 0) [anchor=south] {\( 3 \)};
    
    \addplot[
        only marks,
        mark=*,
        mark size=2pt,
    ] coordinates { (15.7, 0) };
    \node at (axis cs:15.7, 0) [anchor=south] {\( 4 \)};
    
    \addplot[
        only marks,
        mark=*,
        mark size=2pt,
    ] coordinates { (17.3, 0) };
    \node at (axis cs:17.3, 0) [anchor=south] {\( 5 \)};
    
    \addplot[
        only marks,
        mark=*,
        mark size=2pt,
    ] coordinates { (19, 0) };
    \node at (axis cs:19, 0) [anchor=south] {\( 6 \)};
    
    \end{axis}
    \end{tikzpicture}
    \caption{Longitudinal section of wave flume and the location of gauges used for the harmonic wave generation over a submerged bar.}
    \label{fig:bar:test:experiment}
\end{figure}

In this section, we consider a classical benchmark for dispersive and non-linear water waves propagating over a submerged bar and compare the numerical results with experimental data. The test involves uneven bottom topography and poses a challenge for numerical wave models due to the need to accurately capture higher-order harmonics arising from non-linear wave-wave and wave-bottom interactions. The experimental setup of the wave flume is described by the longitudinal section in Figure \ref{fig:bar:test:experiment}. For details on the experimental setup, we refer to Beji et al. \cite{Beji1994}.

We simulate flow over the submerged bar using the numerical discretization described in Section~\ref{sec:discretization}, applied to a structured grid with \( (N_x, N_z) = (612, 9) \) nodes in the \( x \)- and \( z \)-directions. Neumann boundary conditions are imposed on the walls via ghost nodes. Spatial derivatives are computed using sixth-order finite-difference operators, and time integration is performed using the LSERK45 scheme (Algorithm~\ref{alg:time:integration}) with a time step of \( \Delta \tau = 0.025\,\text{s} \) over 3313 steps. At each Runge-Kutta stage, pressure correction involves solving a mixed-stage Poisson problem using GMRES, preconditioned with an incomplete-\( LU \) factorization of the Laplacian at still water level (\( \eta = 0 \)), and solved to a relative tolerance of \( 10^{-4} \). For further detail son the implementation, we refer to Engsig-Karup et al.~\cite{engsig-karup2024}.



\begin{figure}[!h]
    \centering
    \begin{subfigure}[b]{0.45\textwidth}
        \centering
        \includegraphics[width=\textwidth]{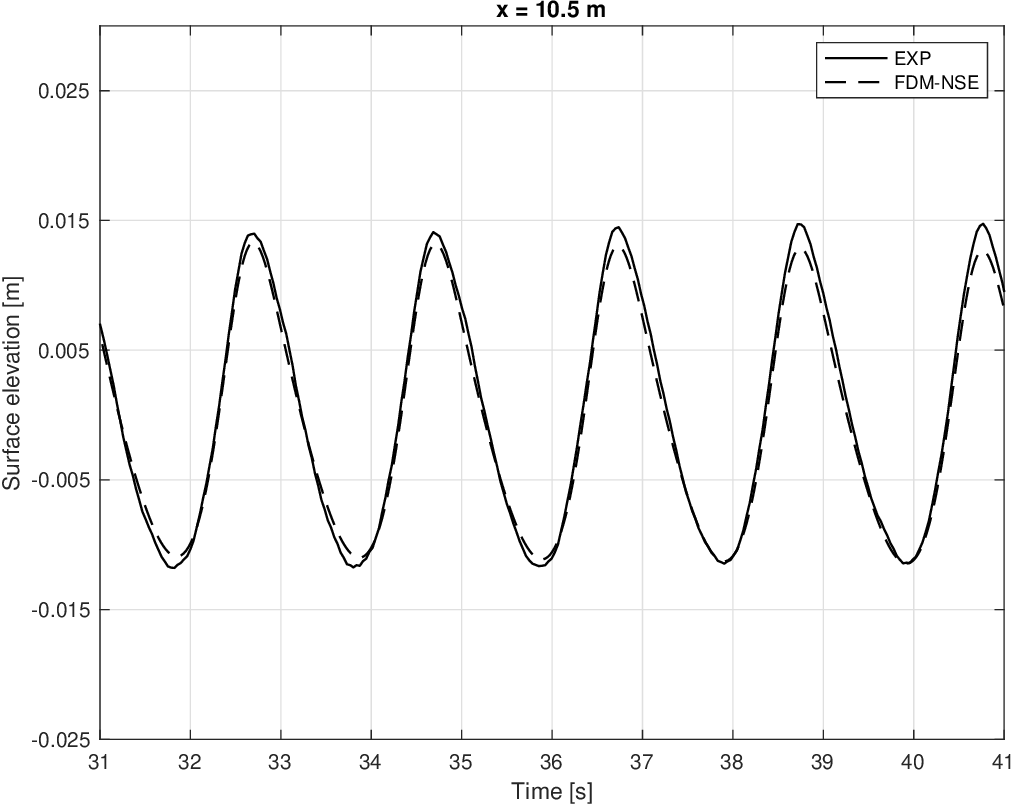}
    \end{subfigure}
    \hspace{1cm}
    \begin{subfigure}[b]{0.45\textwidth}
        \centering
        \includegraphics[width=\textwidth]{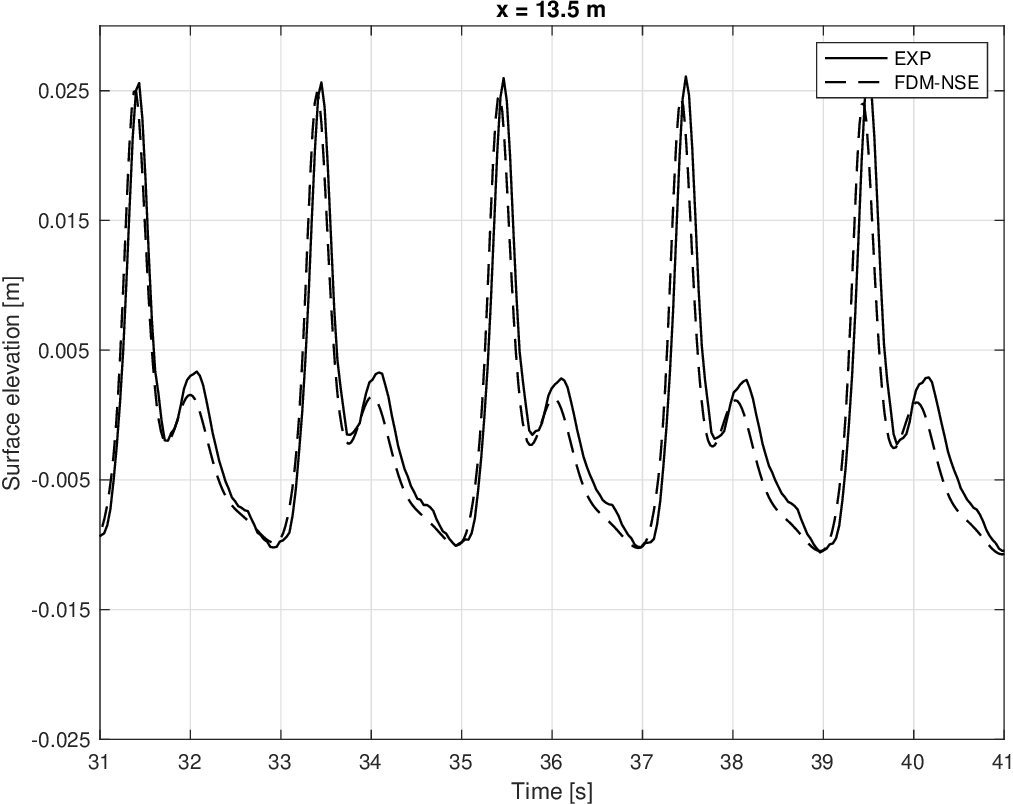}
    \end{subfigure}
    \par\bigskip
    \begin{subfigure}[b]{0.45\textwidth}
         \centering
         \includegraphics[width=\textwidth]{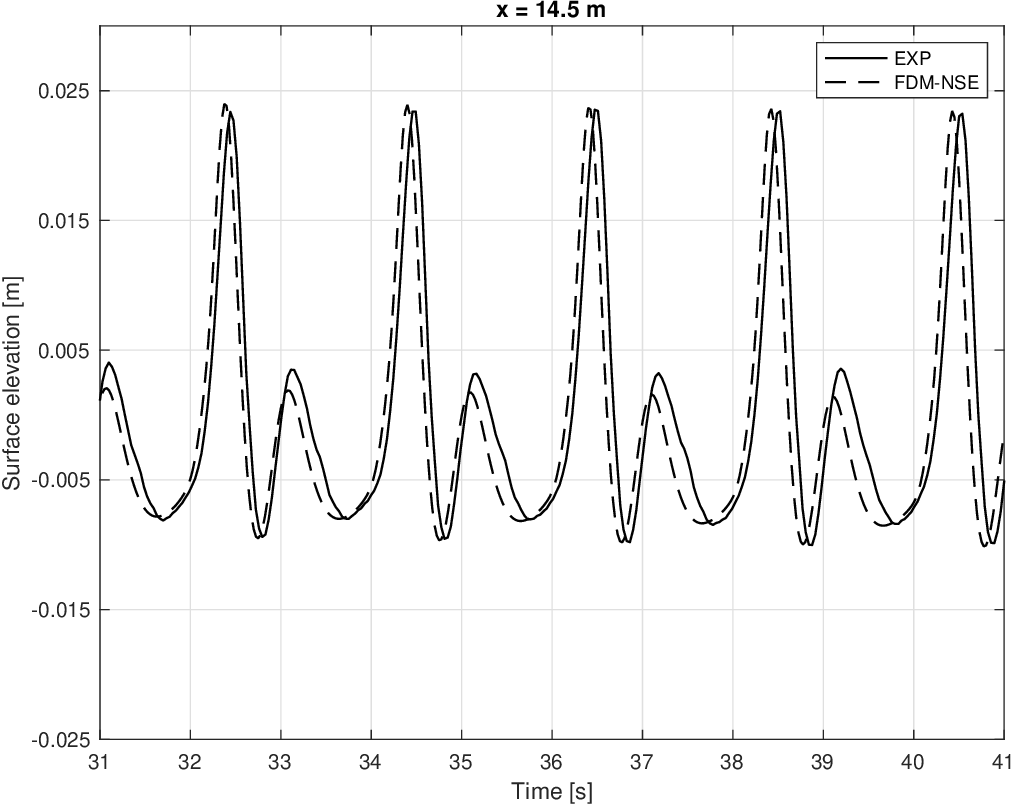}
    \end{subfigure}
    \hspace{1cm}
    \begin{subfigure}[b]{0.45\textwidth}
        \centering
        \includegraphics[width=\textwidth]{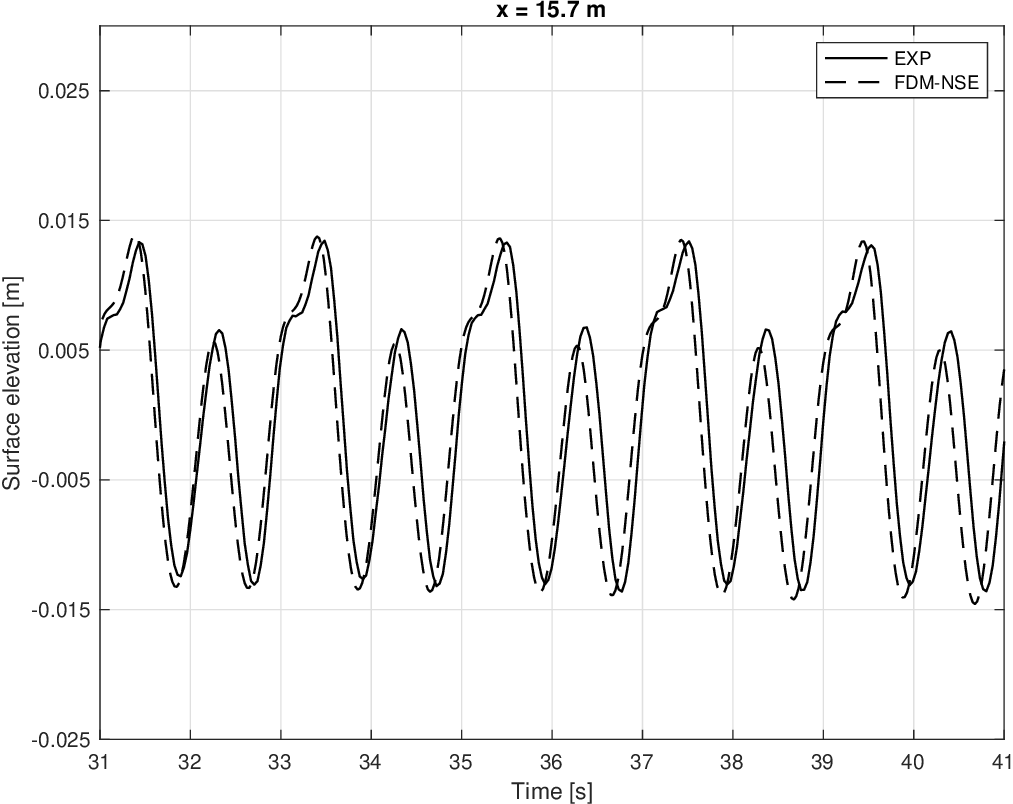}
    \end{subfigure}
    \par\bigskip
    \begin{subfigure}[b]{0.45\textwidth}
        \centering
        \includegraphics[width=\textwidth]{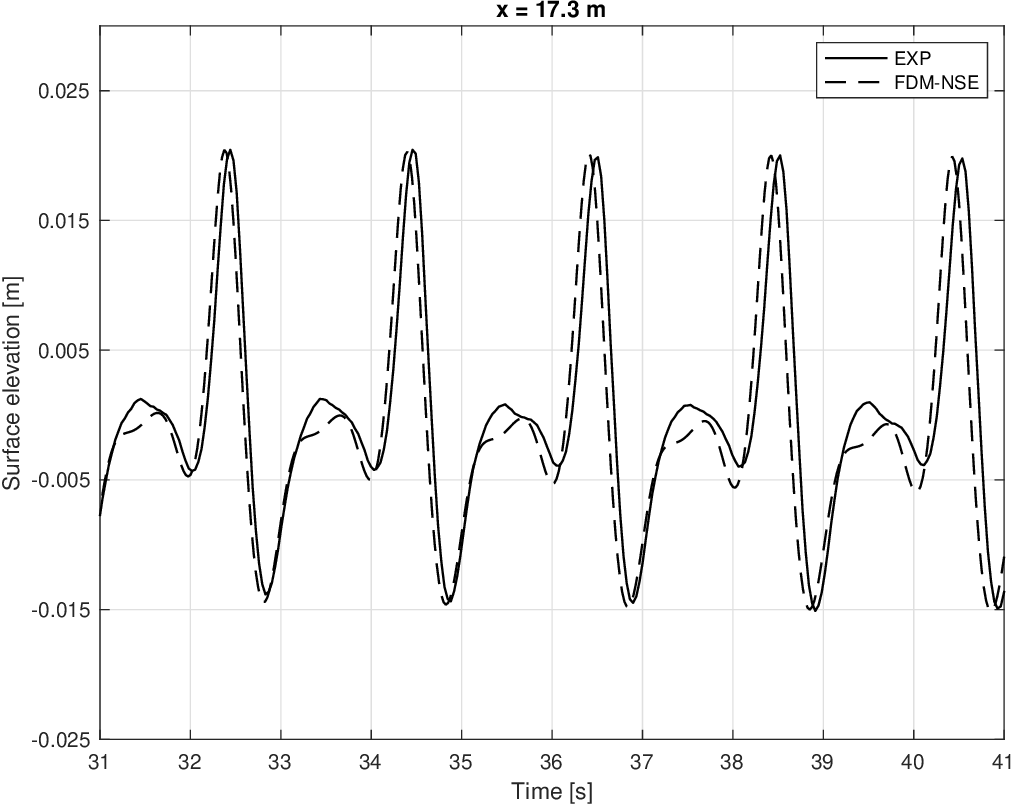}
    \end{subfigure}
    \hspace{1cm}
    \begin{subfigure}[b]{0.45\textwidth}
         \centering
         \includegraphics[width=\textwidth]{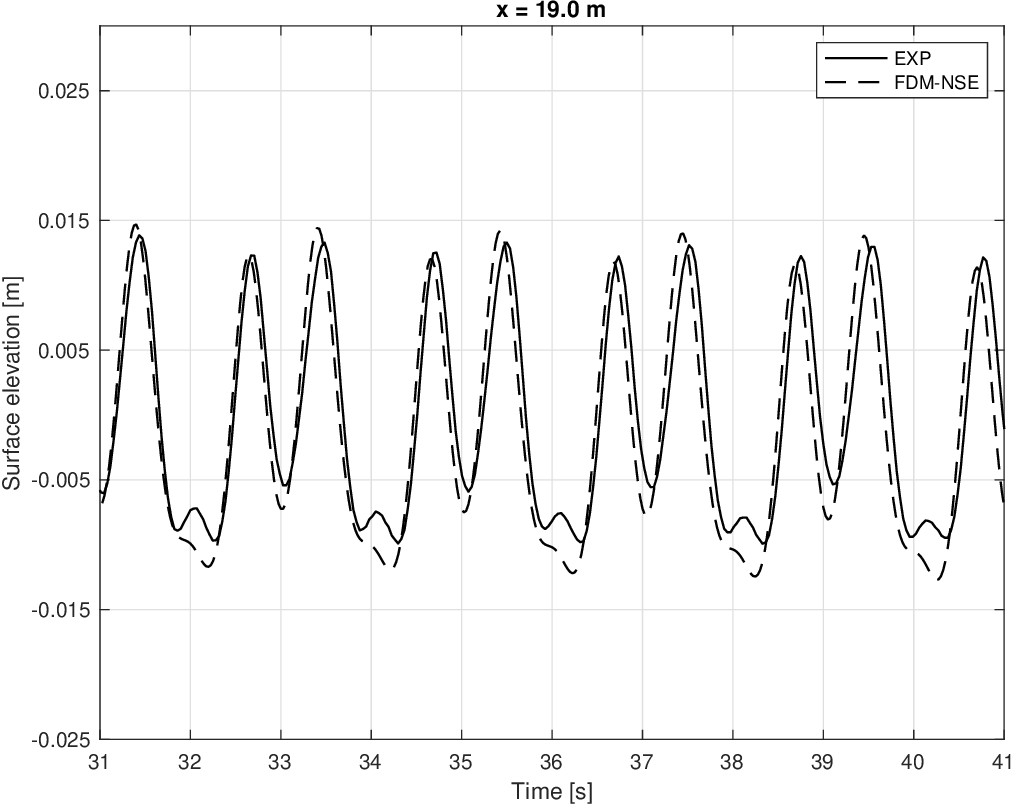}
    \end{subfigure}
    \caption{The figure shows the simulated flow over the submerged bar and experimental data at the gauges.}
    \label{fig:bar:test}
\end{figure}

The simulated surface elevation of the nonlinear and dispersive wave propagation across the submerged bar is compared with experimental measurements at the wave gages, as shown in Figure~\ref{fig:bar:test}. The results demonstrate close agreement between the numerical simulation and the experimental data, confirming the accuracy of the model.


The effect of subspace acceleration using the randomized range finder in Algorithm~\ref{alg:range:finder} is presented in Table~\ref{tab:subspace:dimension:bar:test}. The results show that subspace acceleration reduces the number of GMRES iterations by more than a factor of \( 1.5 \) for most combinations of \( K \) and \( k \). As expected, the average iteration count is similar when constructing the subspace \( \mathcal{S} \) from either the sketch \( X\Omega \) or is \( QR \)-factorization. Minor differences are attributed to numerical round-off errors.

\begin{center}
\begin{table*}[!h]%
\caption{Case: Harmonic wave generation over a submerged bar. The average number of iterations, time and speed up per iteration compared to the baseline using Algorithm \ref{alg:range:finder} for different subspace parameters .}
\label{tab:subspace:dimension:bar:test}
\begin{tabular*}{\textwidth}{@{\extracolsep\fill}cccccccccc@{}}
\toprule
& &\multicolumn{4}{@{}c}{\( V=X\Omega \)} & \multicolumn{4}{@{}c}{\( V=Q \)} \\

\cmidrule{3-6}\cmidrule{7-10}
\( K \) & \( k \) & Iterations & Speed Up  & Time & Speed Up  & Iteration & Speed Up & Time & Speed Up \\
\midrule
10 &  6 & 3.59 & 1.46 & 4.86 & 0.74 & 3.55 & 1.47 & 3.88 & 0.92 \\
10 &  8 & 3.12 & 1.68 & 4.13 & 0.87 & 3.12 & 1.68 & 4.71 & 0.76 \\
15 &  6 & 3.76 & 1.39 & 4.49 & 0.80 & 3.76 & 1.39 & 4.78 & 0.75 \\
15 &  8 & 3.34 & 1.57 & 3.79 & 0.94 & 3.34 & 1.57 & 4.90 & 0.73 \\
15 & 10 & 3.00 & 1.74 & 3.83 & 0.93 & 3.00 & 1.75 & 5.35 & 0.67 \\
20 & 10 & 3.17 & 1.65 & 3.95 & 0.90 & 3.18 & 1.65 & 5.51 & 0.65 \\
20 & 15 & 2.65 & 1.97 & 4.22 & 0.85 & 2.64 & 1.98 & 6.42 & 0.56 \\
30 & 10 & 3.72 & 1.40 & 4.49 & 0.80 & 3.73 & 1.40 & 5.47 & 0.65 \\
30 & 12 & 3.18 & 1.64 & 4.65 & 0.77 & 3.18 & 1.64 & 6.08 & 0.59 \\ 
30 & 15 & 2.91 & 1.80 & 4.51 & 0.79 & 2.91 & 1.79 & 6.86 & 0.52 \\
\bottomrule
\end{tabular*}
\begin{tablenotes}
\item The baseline algorithm uses an average of \( 5.23 \) GMRES iterations and \( 3.58\mathrm{ms} \) per time step.
\end{tablenotes}
\end{table*}
\end{center}

Similar results are observed when applying the exponentially weighted moving average (EWMA). As shown in Table~\ref{tab:forgetful:iterations:bar}, for a fixed subspace size \( k \) and forgetfulness rates \( r \leq 0.8 \), the average number of GMRES iterations per time step is consistently lower compared to the randomized range finder. This supports our earlier findings from the stream function wave case in Table~\ref{tab:iteration:comparison}, indicating that the continuous forgetting mechanism in EWMA leads to fewer iterations for the same subspace dimension \( k \).

\begin{center}
\begin{table*}[!h]%
\caption{Case: Harmonic wave generation over submerged bar. The average number of iterations per time-step different using Algorithm \ref{alg:ewma} for different subspace dimensions and rate of forgetfullness.}
\label{tab:forgetful:iterations:bar}
\begin{tabular*}{\textwidth}{@{\extracolsep\fill}cccccccc@{}}
\toprule
 & 1.0 & 0.95  & 0.9 & 0.85  & 0.8 & 0.75 & 0.7 \\
\midrule
6  & 4.9560 & 4.1196 & 3.8527 & 3.6865 & 3.5917 & 3.5224 & 3.4820 \\
8  & 5.0414 & 3.8430 & 3.4762 & 3.2824 & 3.1785 & 3.1084 & 3.0697 \\
10 & 4.9744 & 3.6094 & 3.1927 & 3.0007 & 2.9060 & 2.8380 & 2.8033 \\
12 & 4.9207 & 3.3955 & 2.9820 & 2.8073 & 2.7053 & 2.6517 & 2.6243 \\
15 & 4.8325 & 3.1714 & 2.7750 & 2.5856 & 2.4744 & 2.4359 & 2.4138 \\
\bottomrule
\end{tabular*}
\begin{tablenotes}
\item The baseline algorithm uses an average of \( 5.23 \) GMRES iterations.
\end{tablenotes}
\end{table*}
\end{center}

Although subspace acceleration significantly reduces the average number of GMRES iterations per time step, we do not  see a corresponding reduction in the average computational time per time step for harmonic wave generation over a submerged bar, as shown in Tables~\ref{tab:subspace:dimension:bar:test} and~\ref{tab:forgetful:time:bar}. This is primarily because our focus has been on minimizing iteration counts rather than optimizing runtime performance. We attribute the lack of time improvement to a suboptimal MATLAB implementation, which does not fully exploit the potential efficiency gains of the reduced iteration counts.

\begin{center}
\begin{table*}[!h]%
\caption{Case: Harmonic wave generation over a submerged bar. The average time in milliseconds per time-step using Algorithm \ref{alg:ewma} for different subspaces dimension and forgetness rates.}
\label{tab:forgetful:time:bar}
\begin{tabular*}{\textwidth}{@{\extracolsep\fill}cccccccc@{}}
\toprule
 & 1.0 & 0.95  & 0.9 & 0.85  & 0.8 & 0.75 & 0.7 \\
\midrule
 6 & 4.2829 & 4.8274 & 4.7584 & 4.7406 & 3.7028 & 3.8927 & 3.6703 \\
 8 & 4.4700 & 4.0053 & 4.7563 & 4.7309 & 3.6963 & 3.7198 & 3.9483 \\
10 & 5.2760 & 4.5586 & 5.2469 & 5.1805 & 4.2683 & 4.0178 & 5.1367 \\
12 & 5.9492 & 5.1454 & 5.4438 & 5.3939 & 4.5690 & 4.2118 & 5.3556 \\
15 & 6.2662 & 5.9635 & 5.9059 & 5.4491 & 4.4485 & 4.4693 & 5.7238 \\
\bottomrule
\end{tabular*}
\begin{tablenotes}
\item The baseline algorithm uses an average \( 3.58\mathrm{ms} \) per time step.
\end{tablenotes}
\end{table*}
\end{center}

\section{Conclusions} \label{sec:conclusion}

In this work, we have demonstrated that subspace acceleration is an effective strategy for substantially reducing the average number of GMRES iterations required to solve the mixed-field Poisson equation for pressure in simulations of nonlinear water waves. Building on the original method proposed by Guido et al.~\cite{guido2024}, we introduced a modification that reduces both computational and memory requirements by avoiding orthonormalization and allowing efficient rank-\( 1 \) updates. Despite its simplicity, our modification not only improves efficiency, but also provides better initial guesses for the linear systems solved at each time step.

We evaluated the proposed strategy using high-order finite difference discretizations for free-surface incompressible Navier-Stokes solutions, as developed by Engsig-Karup et al.~\cite{engsig-karup2024}, and applied it to both stream function wave scenarios and the classical submerged bar benchmark problem. In both cases, we observed a significant reduction in GMRES iterations while maintaining the accuracy of the simulation.

Although our experiments were conducted using finite difference methods, the approach itself is agnostic to the choice of spatial discretization and can be readily integrated into other numerical frameworks. This makes the modified subspace acceleration technique a practical and scalable enhancement for a wide range of nonlinear wave simulation settings.

In ongoing work, the subspace acceleration techniques proposed here are being implemented and tested in massively parallel, large-scale simulation tools, with the aim of further improving performance in high-resolution and high-performance computing environments. More complex experimental setups with wave-structure interactions are also being considered.
    

\section*{Acknowledgments}

This work partly contributes to the activities of the research supported by COWIfonden (Grant no. A-165.19) and granted to Prof. A.P. Engsig-Karup.
This work has been carried out within the framework of the EUROfusion Consortium, via the Euratom Research and Training Programme (Grant Agreement No 101052200 — EUROfusion) and funded by the Swiss State Secretariat for Education, Research and Innovation (SERI). Views and opinions expressed are, however, those of the author(s) only and do not necessarily reflect those of the European Union, the European Commission, or SERI. Neither the European Union nor, the European Commission nor SERI can be held responsible for them.

\section*{Data Availability Statement}
Data supporting the findings of this study are available from the corresponding author upon a reasonable request.

\bibliographystyle{abbrvnat}
\bibliography{references}

\end{document}